\documentclass{emulateapj}
\usepackage{graphicx}   
\usepackage{subfigure}
\usepackage{color}
\newcommand{\Lsun}{L_{\odot}}
\newcommand{\Msun}{M_{\odot}}
\newcommand{\Mvir}{M_{\rm halo}}
\newcommand{\Mstar}{M_\star}
\newcommand{\rhalf}{r_{1/2}}
\newcommand{\Rhalf}{R_{\rm e}}
\newcommand{\kms}{{\rm km}\,{\rm s}^{-1}}

\begin{document}
\title{Stealth Galaxies in the Halo of the Milky Way}
\author{James S. Bullock\altaffilmark{1}, Kyle R. Stewart\altaffilmark{1}, Manoj Kaplinghat\altaffilmark{1}, Erik J. Tollerud\altaffilmark{1}, and Joe Wolf\altaffilmark{1}}

\altaffiltext{1}{Center for Cosmology, Department of Physics and Astronomy, The University of California at Irvine, Irvine, CA, 92697, USA; bullock@uci.edu }

\begin{abstract} {
We predict that there is a population of low-luminosity  dwarf
galaxies orbiting within the halo of the Milky Way 
that have surface brightnesses low  enough to have escaped detection
in star-count surveys.  The overall count of stealth galaxies is
sensitive to the presence (or lack) of a  low-mass threshold in galaxy
formation.   These systems 
have  luminosities and stellar velocity dispersions that are similar
to those of known ultrafaint dwarf  
galaxies but they have more extended stellar distributions  (half
light radii greater than about 100 pc)
because they inhabit dark subhalos that are slightly less massive than
their higher surface brightness counterparts. As a result, the typical
peak surface brightness is fainter than 30 mag per square arcsec. 
One implication is that the inferred common mass scale for Milky Way
dwarfs may be an artifact of selection bias. 
If there is no sharp threshold in galaxy formation at low halo mass,
then ultrafaint galaxies like Segue 1 represent the high-mass, early
forming tail of a much larger population of objects  that could number
in the hundreds and have typical
peak circular velocities of about 8 $\kms$ and masses within 300 pc of
about 5 million solar masses. 
Alternatively, if we impose a low-mass threshold in galaxy formation
in order to explain the unexpectedly high densities of the ultrafaint
dwarfs, then we expect only a handful of stealth galaxies in the halo
of the Milky Way.  
A complete census of these objects  will require deeper sky surveys,
30m-class follow-up telescopes, and more refined methods to identify
extended, self-bound  groupings of stars in the halo.}
   \end{abstract}

\section{Introduction}
\label{Introduction}

Approximately twenty-five new dwarf galaxy companions of the Milky Way (MW) and M31
have been discovered since 2004, more than doubling the known satellite 
population in the Local Group  in five years \cite[][]{willman:05,zuck:06,grillmair:06,maj:07,belo:07,grillmair:09,belok:09,martin:09}.  The majority of these newly-discovered 
dwarfs are less luminous than any galaxy previously known.  The most extreme of these, the ultrafaint MW dwarfs,  
have luminosities smaller than an average globular cluster L$_{V} \simeq 10^2 - 10^4$ L$_\odot$, and were discovered
by searches for stellar overdensities in the wide-field maps of the Sloan Digital Sky Survey (SDSS) and the
Sloan Extension for Galactic Understanding and Exploration (SEGUE).
Follow-up kinematic observations showed that these tiny galaxies have surprisingly high stellar velocity dispersions 
for their luminosities and sizes
 \citep[$\sigma_\star \sim 5 \, \kms$,][]{martin07,sg07,g09} and subsequent mass modeling has shown that
they are the most dark matter dominated galaxies known \citep{Strigari08,wolf:09}.   
Remarkably, these extreme systems are not only the
faintest, most dark matter dominated
galaxies in the universe but they are also the most metal poor  stellar systems yet studied \citep{kirby08,g09}.

Perhaps the most exciting aspect of these recent discoveries is that they point to a much
larger population.  Detection in the SDSS is complete only to $\sim 50$ kpc for  the least luminous
dwarfs \citep{kopo:08,walsh09} and straightforward cosmologically motivated
luminosity bias and coverage corrections suggest that there are between $\sim 200$ and $\sim 500$ 
ultrafaint dwarf galaxies within 400 kpc
 of the Milky Way \citep{Tollerud08}, with an ever increasing number beyond.
Importantly, the luminosity-distance detection limits only apply for systems with peak surface brightness obeying $\mu_V < 30$ mag arcsec$^{-2}$
\citep{kopo:08}. Any satellite galaxy with a luminosity of $L \sim 1000 \, L_\odot$ and a projected half-light radius $\Rhalf$
larger than about 100 pc would have evaded detection with current star-count techniques regardless of its distance from the Sun.

Here we argue that there is likely a population of dwarf galaxies surrounding the Milky Way 
(and by extension, throughout the universe) that are so diffuse in stellar density that they would have thus far 
avoided discovery.  Our predictions rely on the fact that the effective radius $\Rhalf$ of 
a dark matter dominated, dispersion-supported galaxy with fixed stellar velocity dispersion will increase as
its dark matter halo mass decreases.   One implication of this idea is that the known ultrafaint dwarf spheroidals (dSphs) 
may represent the high (dark matter) mass tail of a larger distribution of stealth galaxies.  These undiscovered systems
should preferentially inhabit the smallest dark matter subhalos that host stars (with maximum circular velocities
$V_{\rm   max} \lesssim 15 ~\kms$)
and their possible presence should be accounted for in models that attempt to understand 
the satellite-subhalo problem in Cold Dark Matter (CDM) models \citep{klypin99,moore:99,bkw00,stoehr02,zb03,kravtsov:04,strig:07a,maccio09,busha:09,kravtsov:09,OF09}. 

  The link between subhalo mass and galaxy surface brightness may be particularly important for the quest to identify
   `fossils' of the reionization
epoch in the local universe \citep[e.g.][]{RGS02,RG05,GK06,Orban08,RGS08,madau:08,Bovill_Ricotti2009,munoz:09,Ricotti2010}.  The smallest halos
($V_{\rm   max} \lesssim 15 ~\kms$) are the ones that experience suppression after reionization.  Before reionization,
stars can form in these systems via H$_2$ cooling provided sufficient H$_2$ is available.  Conclusive identification
of these H$_2$-cooling fossils in the Local Group would provide present-day laboratories for studying first-light star formation.
Unfortunately, according to our estimates below, these low-mass, first-light fossils will likely be too diffuse to
discover readily in star-count surveys.  Indeed, \citet{Bovill_Ricotti2009} have shown that low-luminosity fossil dSph galaxies produced in self-consistent cosmological simulations do tend to have very low surface brightness, such that they would not
have been discovered in current surveys.  

Stellar kinematic samples in dSph satellte galaxies
provide a means to directly 
constrain total  dark matter halo masses or $V_{\rm max}$ values on an object by object basis.  
Unfortunately, this
task is not straightforward because dSph dark matter halos are expected to extend well beyond
their stellar radii.  Nevertheless, by  constraining the mass of a galaxy within its
stellar extent, 
one can estimate (for example) a $V_{\rm max}$ value by imposing a prior assumption
that their density profiles behave as predicted for {\em subhalos} in CDM simulations \citep{zb03,Hayashi03,kazantzidis04,penarrubia08}.  
Remarkably, these studies indicate that {\em all} of the known MW dSph satellite galaxies with well-studied
kinematic samples are embedded within massive dark matter subhalos
\citep[$V_{\rm   max} \gtrsim 15 ~\kms$; Wolf et al., in preparation;][]{penarrubia08}, and this includes
ultrafaint dwarfs.  Keep in mind, however, that these determinations rely on significant extrapolations:
the rotation curve of a $V_{\rm max} = 15~\kms$ subhalo typically peaks at 
a radius of $\sim 1700$ pc (Springel et al. 2008; Diemand et al. 2008).  
The median three-dimensional half-light radius for the sample 
of dSphs we consider here (see Wolf et al. 2010, Table 1) is about 300 pc, therefore the extrapolation is fairly large.

Another way to study dSph halo potential well depths is to consider masses within central
radii that are similar in extent to the stars in the galaxies \citep{strig:07b} but still large enough that they can be
resolved directly in numerical simulations \citep{strig:07a,Strigari08}.  Today's state-of-the art N-body simulations
cannot resolve subhalo densities below about 300 pc (Springel et al. 2008; Diemand et al. 2008).  As a fortunate coincidence, 300 pc is also the median half-light radius for the  population of MW dSphs with well-studied kinematic samples
\citep{Strigari08,wolf:09}.   For these practical (not physical) reasons  \citet{Strigari08}  studied the integrated mass  within 300 pc ($M_{300}$) for all of the MW dSphs with large stellar kinematic samples.  They found
$M_{300} \simeq 10^7 \Msun$ and 
no evidence for a relationship between halo $M_{300}$ and
 total luminosity:  $M_{300} \propto L^{0.03 \pm 0.03}$. 
The implication is that the tiniest  
satellite galaxies have dark matter densities indicative of dark matter halos that are
at least as massive as those of systems 10,000 times more luminous.

The normalization of the Strigari et al.  relation  at
  $M_{300} \simeq 10^7 \Msun$ is indicative of
  the central densities of {\em massive} dark matter subhalos ($V_{\rm
    max} \gtrsim 15 ~\kms$).  As alluded to above, this mass-scale is fairly easy to explain in $\Lambda$CDM models
  \citep{Strigari08,maccio09,busha:09,munoz:09,kravtsov:09,Li09,OF09,Stringer09}.
However, the lack of observed correlation between $L$ and mass is
quite unexpected.  To put the lack of measured slope in 
the $M_{300} - L$ relation in perspective, consider the relationship
between dark matter halo $V_{\rm max}$ and galaxy luminosity 
required to match the faint-end slope of the galaxy luminosity function:
$L \propto V_{\rm max}^{b}$ with $b = 7.1$ \citep{busha:09}. Tully-Fisher studies
suggest a relation that is even less steep for more massive systems, with $b \simeq 4$ \citep[e.g.,][]{Stark09,Courteau07,MW10}.
For dark
matter halos of interest, with maximum circular velocities 
$V_{\rm max} \simeq 15 - 45 \, \kms$, we expect $M_{300} \propto
V_{\rm max}$ \citep[assuming NFW fits to halos in][]{springel:08}
such that the observed trend $M_{300} \propto L^{0.03}$ would naively
imply $L \propto V_{\rm max}^{b}$ with $b \simeq 33$.  This is a much steeper relationship
than we expect from more luminous systems. 
The lack of inferred relationship
between dSph luminosity and total halo mass is not an artifact of the specific choice of
300 pc for the mass comparison.   If one performs the same comparison between
galaxies using a smaller benchmark radius \citep[100
pc,][]{Strigari08} or using the 3d half-light radius for each galaxy
\citep{wolf:09} or using the 3d mass within the 2d half-light radius
of each galaxy \citep{walker:09}   then one reaches the same
conclusion: there is no observed trend between inferred {\em total} halo mass (or
$V_{\rm max}$) and luminosity.~\footnote{Note that there {\em is} an observed trend between
the mass within the {\em half-light radius} of each galaxy and the galaxy's luminosity, 
but this simply reflects the fact that brighter galaxies have larger half-light radii.  This 
trend is perfectly consistent with each of the galaxies being embedded within halos
of approximately the same total mass \citep{walker:09,wolf:09}.}

One possible
explanation for this lack of mass trend is that it reflects a scale in
galaxy formation, where the scatter in $L$ at fixed $V_{\rm max}$  
becomes very large, as might possibly be explained 
by feedback due to photoionization
  or  suppression below the atomic cooling limit 
 \citep[][and \S 3 below]{RGS08,Strigari08,maccio09,OF09,Stringer09}. 
Another possibility, outlined below, is that the lack of an observed
trend between mass and luminosity is the product of selection bias: 
most ultrafaint galaxies do inhabit halos with $M_{300} \lesssim 10^7
\Msun$, but they are too diffuse to have been discovered. 

%>>>>>>>>>>>>>>>>>>>>>>>>>>>>>>>>>Empirical Plots<<<<<<<<<<<<<<<<<<<<<<<<<<<<<<<<<
\begin{figure*}
   \includegraphics[width=0.49\textwidth]{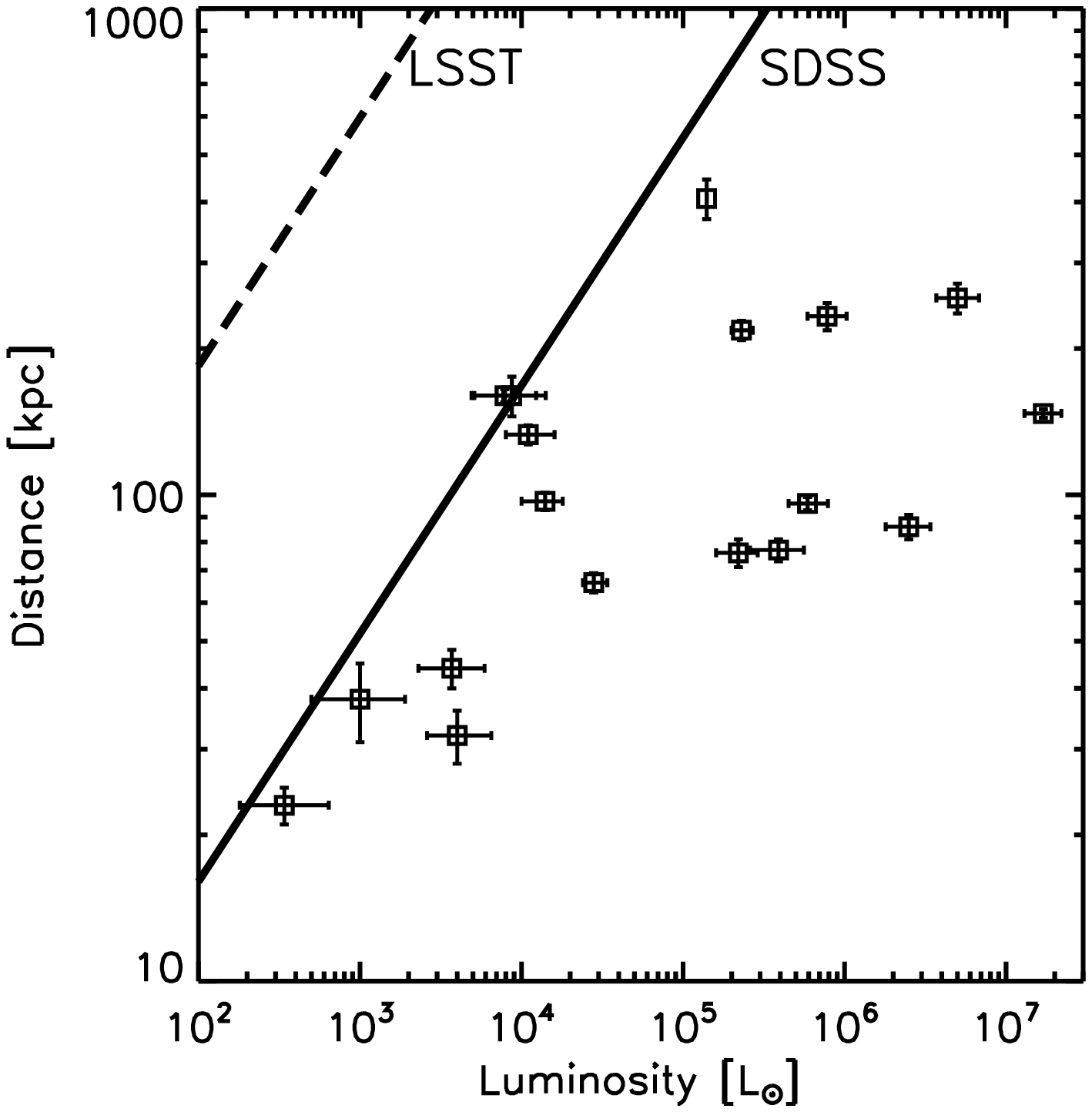}
    \includegraphics[width=0.49\textwidth]{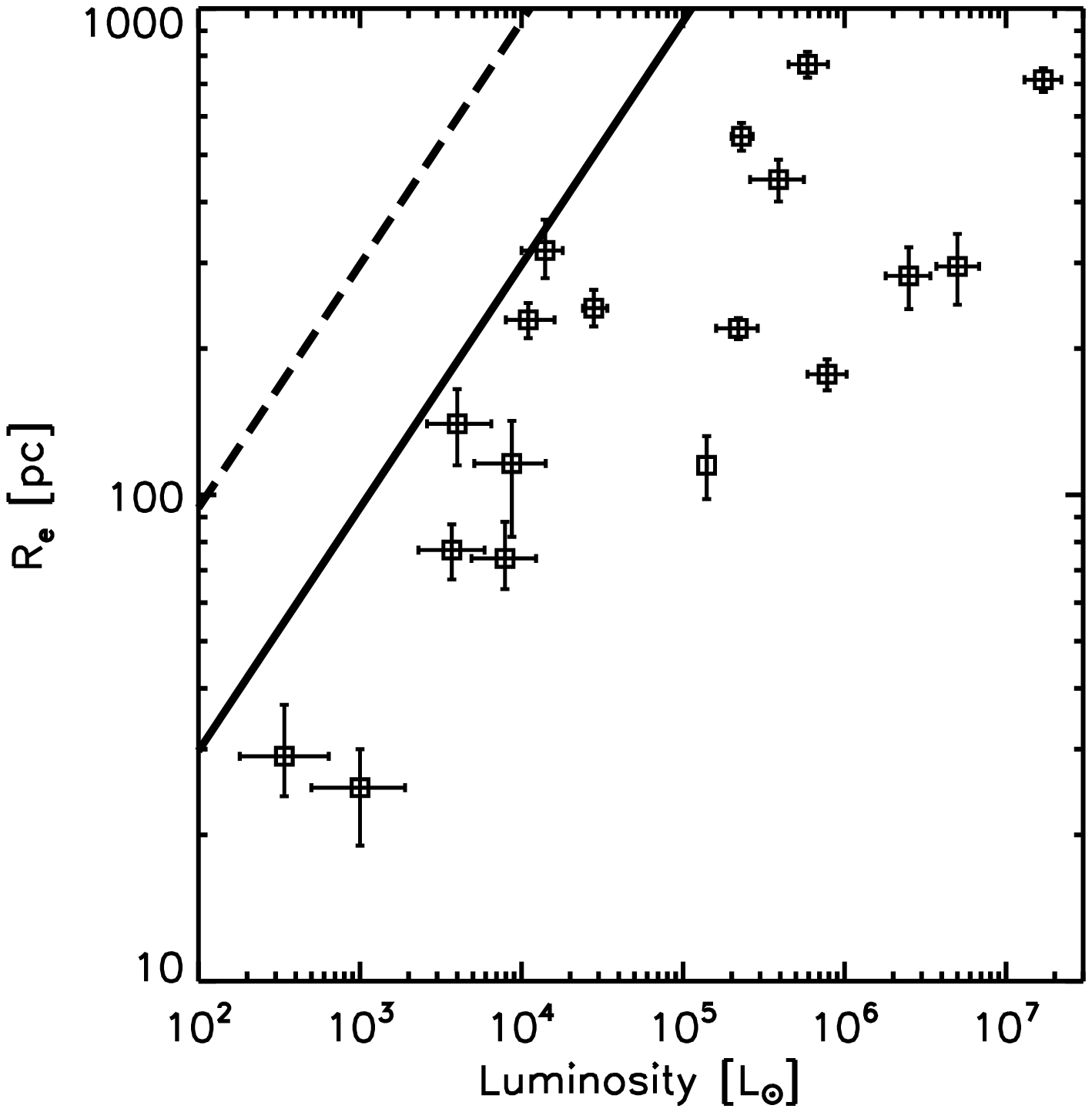}
   \caption{   Projected  helio-centric distance vs. V=band luminosity (left) and
   half-light radius $R_e$ vs. V-band luminosity (right) and
for Milky Way dSph galaxies.
The lines in the left panel show SDSS luminosity completeness limits
from Walsh et al. (2009) and an estimate for LSST completeness following Tollerud et al. (2008).  In the right
panel, galaxies above the solid line, with surface brightness fainter than $\mu = 30$
mag arcsec$^{-2}$ are currently undetectable.  For reference, the dashed line in the right panel corresponds to $\mu = 35$ mag arcsec$^{-2}$. }
\label{empirical1}
\end{figure*}
%>>>>>>>>>>>>>>>>>>>>>>>>>>>>>>>>>>>>>>>><<<<<<<<<<<<<<<<<<<<<<<<<<<<<<<<<

In the next section, we explain why we expect surface brightness selection bias to limit the discovery of
satellite galaxies in small subhalos.  In \S 3 we use a simple model
to estimate the number of low surface brightness stealth  galaxies within the vicinity of the Milky Way.  Our estimates
rely on the
public subhalo catalogs provided by the Via Lactea 2 (VL2) N-body simulation group \citep{VL2}.  We explore
two models. Our Fiducial Scenario
connects each subhalo's mass and accretion time to 
a galaxy luminosity $L$ by extrapolating the
halo mass-light relationship required to match the asymptotic slope of the galaxy stellar mass function  \citep[][]{Moster09}.  
Our secondary model (Threshold Scenario) explores a scenario where galaxy formation is truncated sharply below a characteristic 
dark halo mass scale.
We present our findings in \S 4 and conclude in \S 5.

One of the goals of our model is to investigate how surface brightness detection limits can affect our interpretation
of dwarf subhalo mass-luminosity trends.  We will use $M_{300}$ as our primary mass variable for characterizing
subhalo masses.  As discussed earlier, $M_{300}$ can be measured directly for the N-body subhalos, but 
requires less extrapolation than $V_{\rm max}$ when comparing to observational data.  
For the observational comparisons, we will use $M_{300}$ values derived by Strigari et al. (2008) for MW dwarfs.
We emphasize that the choice of 300 pc is a practical one, with no special physical meaning, 
other than 300 pc is the median half-light radius of the dSph galaxies in our sample.  Integrated masses within smaller radii
cannot be resolved
in the simulation.  Eight of the nineteen galaxies we consider have 3d half-light radii smaller than 290 pc and there are four galaxies that do not have at least one kinematic stellar tracer beyond this radius.  In these cases we are relying on the CDM-motivated prior of
Strigari et al. (2008) to extrapolate masses out to 300 pc.  Such an extrapolation is perfectly reasonable (and inevitable because of resolution) as long as our aim is to compare to predictions from {\em the same theory} that motivates the prior (as is the case here).Ê The assigned error bars on the measured $M_{300}$  take into account uncertainties in the extrapolation encompassed by the theory, including an allowance for exponential mass truncation because of tides.  
We note that the instantaneous tidal radius of the closest dSph, Segue 1, is much larger than 300 pc (Geha et al. 2009).  Moreover, by examining the orbits of subhalos in VL2, Rocha et al. (2010, in preparation) find that
subhalos chosen to have radii and masses consistent with Segue 1, have past orbital trajectories and measured tidal radii that are larger than $300$ pc in the vast majority of cases.

%>>>>>>>>>>>>>>>>>>>>>>>>>>>>>>>>>Empirical Plots<<<<<<<<<<<<<<<<<<<<<<<<<<<<<<<<<
\begin{figure*}[t!]
  \includegraphics[width=0.49\textwidth]{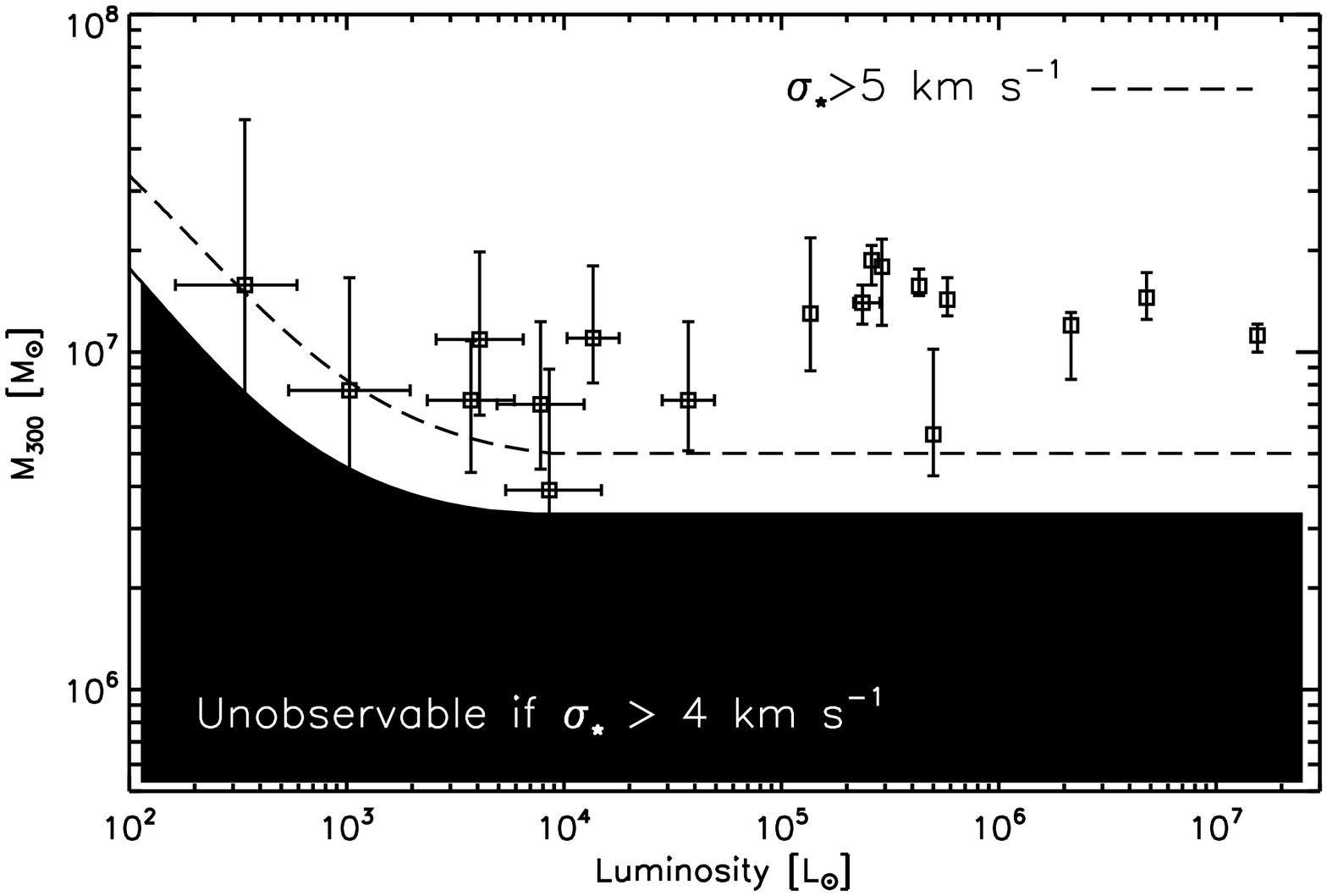}
  \includegraphics[width=0.49\textwidth]{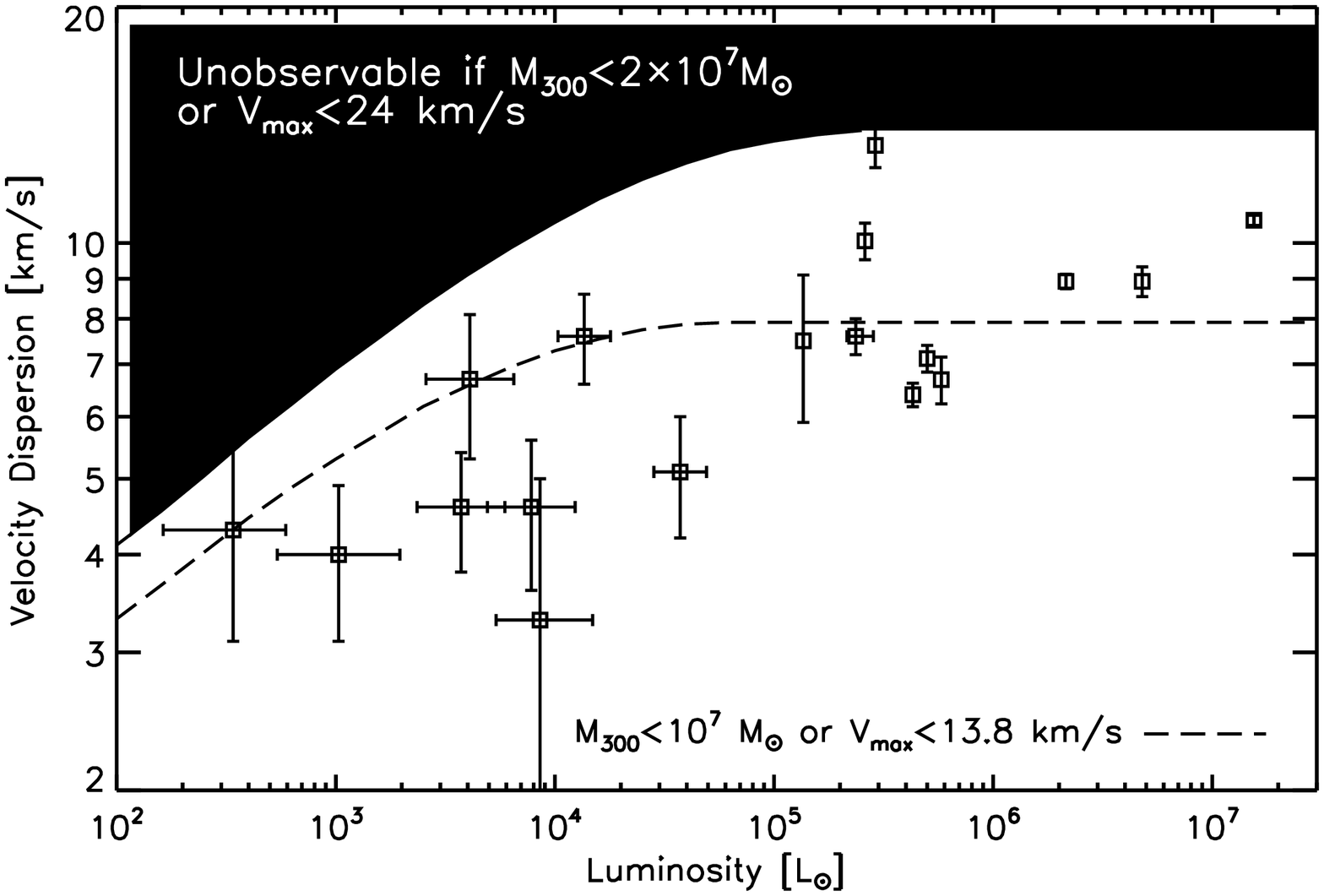}
   \caption{ Mass within 300 pc vs. luminosity (left) and measured stellar velocity velocity dispersion, $\sigma_\star$, vs. luminosity (right)
    for Milky Way dwarf spheroidal galaxies (data points with error bars).  In the left panel, galaxies within in the shaded region (below dashed line) will remain hidden ($\mu > 30$ mag arcsec$^{-2}$) 
    if they have $\sigma_\star > \, 4 \kms$ ( $\sigma_\star > 5 \, \kms$).  In the right panel, galaxies in the shaded region will be stealth ($\mu > 30$) if they have $M_{300} < 2 \times 10^7 M_\odot$ and galaxies below the dashed line will be stealth if they have $M_{300} < 10^7 M_\odot$.  The masses and velocity dispersions quoted here are taken from \cite{Strigari08} and \cite{wolf:09}.  Since the time of those publications, it has come
    to our attention that the velocity dispersion (and hence mass) errors on Hercules \citep[at $L \simeq 3 \times 10^4 L_\odot$,][]{Aden09} and W1 (at $L \simeq 10^3 L_\odot$, B. Willman \& M. Geha, private communications) 
    are likely underestimated because of membership uncertainties. }
\label{empirical2}
\end{figure*}
%>>>>>>>>>>>>>>>>>>>>>>>>>>>>>>>>>>>>>>>><<<<<<<<<<<<<<<<<<<<<<<<<<<<<<<<<

\section{Motivations}

The left panel of Figure 1 shows the MW dSphs as tabulated in \citet{wolf:09} 
plotted in the plane of helio-centric distance vs. V-band luminosity.
The solid line (labeled SDSS)  illustrates the distance to which dwarfs of a given luminosity can be detected in SDSS
with 90\% efficiency from \citet{walsh09}.  Similar results were presented by \citet{kopo:08}.  
The upper line shows the same limit adjusted up by scaling to the limiting magnitude of the full co-added LSST survey
\citep{Tollerud08,Ivezic08}. Clearly, the known dwarf galaxies cluster at the current completeness edge of the diagram, 
indicating a high likelihood for future discoveries \citep{kopo:08,Tollerud08,walsh09}.  

The distance-luminosity completeness limits presented by Walsh et al. (2009) and \citet{kopo:08}  are only
applicable for systems with surface brightness brighter than $\mu_V = 30$ mag arcsec$^{-2}$ \citep[][and G. Gilmore, M. Geha, and B. Willman, private communications]{kopo:08}.  Systems more diffuse than this limit cannot be detected in SDSS with current methods, no matter their
helio-centric distance. This phenomenon is illustrated qualitatively 
in the right panel of Figure 1, which presents the same set of MW dSphs in the plane of
$\Rhalf$ vs. $L$.  
The solid line shows a constant peak (central) surface brightness for a Plummer profile
\begin{equation}
 \Sigma_{\rm peak} = \frac{L}{\pi \Rhalf^2} = 0.036 \, L_{\odot} \, {\rm pc}^{-2},
 \label{eq:plum}
 \end{equation}
and corresponds to $\mu_V = 30$ mag arcsec$^{-2}$ for solar absolute magnitude $M_{\odot V} = 4.83$.
As in the distance-luminosity figure, the tendency for many of the fainter dwarfs to ``pile up" 
near the surface brightness  detection limit is suggestive.  There is nothing ruling out the presence of a larger population of more extended systems that
remain undetected because of their low surface brightness.  

If a large number of diffuse, undetected galaxies do exist, they are likely associated with low-mass dark matter subhalos.
One can understand this expectation by considering
an spherically-symmetric galaxy that is in equilibrium 
with stellar density distribution $\rho_*(r)$ and radial velocity profile $\sigma_r(r)$ that is
embedded within a gravitationally-dominant dark
matter halo mass profile $M(r)$.  The Jeans equation is conveniently written as
\begin{equation}
M(r) = \frac{r \: \sigma_r^2}{G} \left (\gamma_\star+\gamma_\sigma - 2\beta \right ),
\end{equation}
where $\beta(r) \equiv 1- \sigma_t^2 / \sigma_r^2$ characterizes the tangential velocity dispersion and 
$\gamma_{\star} \equiv - {\rm d}\ln \rho_\star / {\rm d} \ln r$ and $\gamma_{\sigma} \equiv - {\rm d}\ln \sigma_r^2 / {\rm d} \ln r$. 
If we make the simplifying assumption that $\beta=0$ and $\sigma_r(r) \simeq \sigma_\star = $ constant, with $\gamma_\sigma \ll 1$ then
 $M(r) = r \, G^{-1} \, \sigma_\star^2 \, \gamma_\star$.  For a fixed
velocity dispersion, a more spatially extended profile (smaller $\gamma_\star$) requires
a lower mass at fixed radius.

The same basic expectation follows in a more general context 
from the recent work of Wolf et al. (2010), who
 showed~\footnote{Under the assumption that the observed stellar velocity dispersion remains fairly flat with projected radius, as is the case with all of the well-studied
 systems.}  that the total mass of a quasi-spherical dSph galaxy within its 3d half-light radius $\rhalf \simeq 1.3 \, \Rhalf$
 may be determined accurately 
from the luminosity weighted line-of-sight velocity dispersion $\sigma_\star$ for general $\beta$: $M(\rhalf) = 3 \, G^{-1} \, r_{1/2} \, \sigma_\star^2$.
Mass determinations at larger and smaller radii require an extrapolation of the mass profile from that point, but given a theoretical
prediction for the mass profile shape $M(r)$  one can perform this extrapolation by simply normalizing at $r=r_{1/2}$.

It is useful to rewrite the Wolf et al. (2010) mass estimator in terms of the implied  circular velocity at $\rhalf$:
\begin{equation}
\label{eq:joewolf}
V_c(\rhalf) = \sqrt{3} \, \sigma_\star.  
\end{equation}
Consider then a galaxy with velocity dispersion $\sigma_\star$ and luminosity $L$
embedded within a gravitationally-dominant dark matter halo described by a
circular velocity curve that increases with radius as an approximate power law:
$V_c(r) = V_{300} \, (r/300 \, {\rm pc})^\alpha$. 
 Equation \ref{eq:joewolf} implies $\rhalf = 300 \, {\rm pc} \, (\sqrt{3} \sigma_\star/ V_{300})^{1/\alpha}$.
 As an example, for an NFW halo \citep{nfw} with $r_s \gg 300~ {\rm pc}$ we have $\alpha = 1/2$ and 
 $\rhalf \propto V_{300}^{-2} \propto M_{300}^{-1}$.    Clearly, the galaxy becomes puffier as we decrease $M_{300}$ or $V_{300}$.
One implication is that if a galaxy has a stellar density that is just large enough to be detected,  
another galaxy with identical $L$ and $\sigma_\star$ will be undetectable if it happens to reside within a slightly less massive halo.

Figure 2 provides a more detailed exploration of the relationship between halo mass parameters ($M_{300}$ or $V_{\rm max}$) and
associated dSph observables  $\sigma_\star$, $L$, and $\Rhalf$. 
Points in the left panel of Figure 2 present  $M_{300}$ vs. $L$ for MW dSph galaxies, with masses from \citet{Strigari08} and luminosities
updated as in \citet{wolf:09}.  The right panel shows $\sigma_\star$ vs. $L$ for the same galaxies culled from Table 1 of \citet{wolf:09}.  

The shaded bands and dashed lines in each panel of Figure 2 illustrate the way in which 
surface brightness incompleteness may affect these diagrams.  In determining these regions we have 
assumed each dSph is dark-matter dominated, such that its gravitating mass profile produces an NFW circular velocity curve $V_c(r) = V_{\rm NFW}(r)$.
Given the NFW shape, the 
rotation curve is fully specified by its peak value $V_{\rm max}$ and the radius where the peak occurs $r_{\rm max}$
\citep[e.g.,][]{b01}.  
We assume for simplicity that subhalos of a given $V_{\rm max}$ map in a one-to-one way to a rotation curve shape using
 $r_{\rm max} = 650 \, {\rm pc} \, (V_{\rm max}/10 \kms)^{1.35}$, which is indicative of median subhalos in high-resolution
N-body simulations (intermediate between the normalizations of Springel et al. 2008 and Diemand et al. 2008).
With this assumption in place, given a halo mass variable (e.g., $M_{300}$ or $V_{\rm max}$),
we may determine
 the implied half-light radius  $\Rhalf \simeq 0.75 \, \rhalf$ associated with any $\sigma_\star$
using $V_{\rm NFW}(\rhalf) = \sqrt{3} \, \sigma_\star$ (Equation \ref{eq:joewolf}).

In the right panel of Figure 2, galaxies
residing in the shaded region are unobservable if they sit within dark matter halos less massive than $M_{300} = 2 \times 10^7$ or
(equivalently for our assumptions) with peak circular velocity smaller than $V_{\rm max} = 24 \, \kms$.  Similarly, galaxies residing above the dashed line are too diffuse to be detected if
they have $M_{300} < 10^7 \Msun$ or $V_{\rm max} \lesssim 14 \, \kms$.  Galaxies need to have deep potential wells if they are
to remain observable at low luminosity for $\sigma_\star \sim 5 \, \kms$.  
If there are low-luminosity galaxies with $M_{300}$ values
smaller than $\sim 10^7 \Msun$ they would remain hidden as long as they have stellar velocity dispersions comparable to those
of the known dwarfs.~\footnote{In deriving these regions, we have explicitly assumed that the stellar systems are dark-matter dominated
within their half-light radii.  The same arguments cannot be applied to globular cluster systems, some of which do inhabit the shaded regions
in the right panel of Figure 2 without any discernible dark matter halo.  These systems have large velocity 
dispersions simply because they have very high stellar densities.}

A related set of limits in the $M_{300}$ -- $L$ plane is depicted in the left panel of Figure 2.
Galaxies sitting in the
shaded band will have $\Rhalf$ too large to
be observable if they have $\sigma_\star > 4 \, \kms$.  Slightly hotter
galaxies, with $\sigma_\star > 5 \, \kms$ will be unobservable if they sit below the dashed line.  As expected, the hotter the galaxy,
the deeper the potential well needs to be in order keep the stars confined to an observable surface brightness.
We see that galaxies with $L \sim 10^3 \, \Lsun$ 
residing in a halos less massive than $M_{300} \simeq 8 \times 10^6 \, \Msun$ will 
be too diffuse to be seen if they have $\sigma_\star = 5 \, \kms$.  
Note that 
for $L \gtrsim 10^4 \Lsun$ the constraint on allowed $M_{300}$ values is flat with $L$ because halos smaller than this
value are kinematically forbidden via Equation \ref{eq:joewolf}.  Specifically, kinematic mass determinations
 demand $V_{\rm max} \ge \sqrt{3} \, \sigma_\star$.

Implicit in the above discussion is the idea that a galaxy's $\sigma_\star$ can be considered independently of its halo mass.  Dynamically,
the only constraint is that $\sigma_\star \le V_{\rm max} / \sqrt{3}$ (Equation \ref{eq:joewolf}).  One is more inclined to suspect that $\sigma_\star$ in an ultrafaint dSph is governed by star formation and galaxy formation processes, with an absolute minimum set by the effective temperature of the star forming ISM.  Even for a very cold primordial effective ISM temperature, $T_{ISM} \sim 300$K, we expect $ \sigma_{ISM} \sim 2 \, \kms$, and this ignores  turbulent and magnetic pressure terms.  
The vast majority of globular clusters have stellar velocity dispersions larger than this \citep{Pryor93}.
Moreover, dark matter halos of all masses are expected to have experienced significant mergers in their early histories \citep[e.g.,][]{Stewart08}.  These mergers would have heated (the oldest) stars beyond any primordial pressurized motions,
and this effect is indeed seen in cosmological simulations of dwarf galaxy formation \citep{Ricotti2010}.

In the next section we consider the implications of a model where $\sigma_\star$ is correlated with luminosity $L$ in a way that tracks the observed
relationship (right panel of Figure 2). In principle, there could be a floor in the $\sigma_\star$ values allowed for dwarf galaxies.  We do not impose such a floor in our calculations, but if one does exist, then our estimates could
{\em under-predict} the fraction of stealth galaxies at low luminosities.

%>>>>>>>>>>>>>>>>>>>>>>>>>>>>>>>>>Moster Lum func<<<<<<<<<<<<<<<<<<<<<<<<<<<<<<<<<
\begin{figure}[t!]
\begin{center}
    \includegraphics[width=0.49\textwidth]{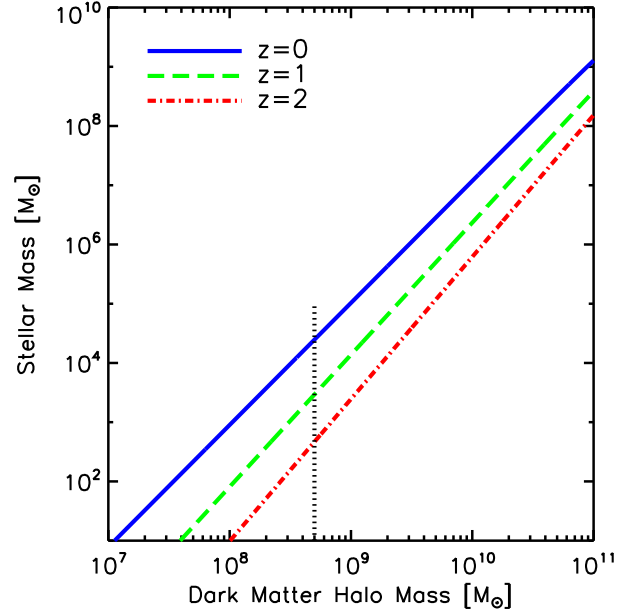}
\end{center}
   \caption{Model stellar mass - halo mass relation \citep{Moster09} shown at three example redshifts.  
   Our Fiducial Scenario assumes that the $\Mvir - M_\star$ relation extrapolates smoothly to very small masses.  
   Our Threshold Scenario imposes a sharp truncation mass at  $\Mvir = 5 \times 10^{8} \Msun$ (vertical dotted line)
    below which all halos are assumed to form no stars.   
   }
\label{mapping}
\end{figure}
%>>>>>>>>>>>>>>>>>>>>>>>>>>>>>>>>>>>>>>>><<<<<<<<<<<<<<<<<<<<<<<<<<<<<<<<<

\section{Model}

We rely on the publicly released subhalo catalogs of the Via Lactea II N-body simulation (VL2 hereafter)
as described in \cite{VL2}.  The simulation adopts cosmological
parameters from WMAP3 \citep{WMAP}  and tracks the formation of a Milky Way size dark matter halo
with a highest particle-mass resolution of $4,100 M_{\odot}$ and force
resolution of $40$ pc.  
The main halo has a radius of 
$402$ kpc, defined to enclose a mean density that is $200$ times the mean matter density of the universe,
and an associated mass of $\Mvir = 1.9 \times 10^{12} M_{\odot}$.  
The public subhalo catalogs include  $M_{300}$, $V_{\rm max}$, and $r_{\rm max}$ parameters for each bound system,
as well as merger history information that allows us to track the redshift of infall $z_{\rm inf}$ for each subhalo and
to determine its maximum attained mass $M_{\rm max}$ prior to infall.
Tests by the VL2 team suggest that the measured $M_{300}$ masses are good to about $20 \%$ (random) owing to
resolution effects (J. Diemand, private communication).

We assign light to each of our accreted dark matter subhalos by assuming that at each redshift $z$ there is
a monotonic relationship between halo mass $\Mvir$ and galaxy stellar mass $M_\star$.  This general approach is
motivated by its demonstrated success in producing the clustering properties of 
galaxies larger than $M_\star \simeq 10^9 \Msun$
 \citep{Kravtsov04a, Tasitsiomi04, ValeOstriker04,
Conroy06, Berrier06, Purcell07, Marin08, Stewart08b,cw09}.  Of course, cosmological abundance
matching cannot be applied directly
at the  smallest stellar masses because of completeness issues.
In our Fiducial Scenario we simply adopt the asymptotic $\Mvir$ -- $M_\star$
relationship suggested by the most complete stellar mass functions, which effectively assumes that there is
no new (or abrupt) mass scale that truncates galaxy formation in small halos.  We also explore a
Threshold Scenario that imposes such a truncation scale (see below).

For our Fiducial Scenario, we assign $M_\star$ to each subhalo by 
extrapolating the fitting formula presented by \citet{Moster09} to small stellar masses.   Moster et al. (2009) derived the relationship using
N-body halo catalogs together with
observationally inferred stellar mass functions for $M_\star \gtrsim 10^{8-9} \Msun$ galaxies out to redshift $z \sim 3$.  
The implied (extrapolated) relationship between stellar mass and dark halo mass is presented in 
 Figure 3 for three example redshifts.  We see that $M_\star$ must decrease at high redshift for a fixed $\Mvir$ in order to explain
the evolving stellar mass function.  Low-mass halos at high redshift have not had time to form as many stars as their
$z \sim 0$ counterparts.  For our Threshold Scenario we adopt the same 
mapping for massive halos but we
impose a sharp truncation in the $M_\star$ - $\Mvir$ relation at $\Mvir = 5 \times 10^8 \Msun$
(dotted line in Figure 3).

We assume that star formation is quenched in each subhalo
at a time $\tau_{\rm q}$ after the redshift of accretion into the VL2 host.  Specifically,
subhalo light content is determined at redshift $z_q$ (set at a time $\tau_q$ after the accretion redshift) 
using  the appropriate Moster et al. (2009) mapping with $M=M_{\rm max}$, the maximum mass each subhalo progenitor obtained prior to infall.    If the subhalo is accreted at a time less than $\tau_{\rm q}$
 before $z=0$ we adopt the Moster et al. (2009) relation at $z_{\rm q} = 0$.
For the figures we present below we use a quenching timescale that is roughly a dynamical time for the host halo 
$\tau_{\rm q} = 2$ Gyr.  We find that the value of $\tau_{\rm q}$ only affects our predictions for the largest satellites
$M_\star \gtrsim 10^6 \, M_\sun$.  For example, if we set $\tau_{\rm q} = 0$, we under-predict the number of luminous
satellites by a few, but the low-luminosity satellite count is largely unaffected.  
The main conclusions of this paper regarding the least luminous, stealth satellites are not sensitive to the
choice of $\tau_{\rm q}$.
For all satellites, we convert from stellar mass to V-band luminosity using
$\Mstar/L = 2 \, [\Msun/\Lsun]$, which is typical for Milky Way dSphs according to \citet{Martin08} for a Kroupa IMF.

In assigning luminosities to dark matter subhalos at the time of infall, we are effectively assuming
that the majority of surviving dark matter halos in our model have lost very little {\em stellar} mass after infall
(though almost all of them lose dark matter mass).   We do make a crude self-consistency check
for stellar tidal mass loss below, but find that it is not significant in most cases.
The idea that stellar mass loss has been minimal for dense, bound dark substructures in the halo
is difficult to test empirically with existing dwarf data \citep[e.g.,][]{Pen09a}.  
Therefore, in order to check whether this assumption is reasonable (at least in the context of the $\Lambda$CDM-based
model we explore here), we examined the output the \citet{BJ05} simulations,
which modeled hundreds of satellite accretions each for
 11 $\Lambda$CDM merger histories using  cosmologically-derived accretion times and orbits,  set within an
 evolving Milky-Way-like disk and dark halo potential. We found that 
 the overwhelming majority of satellite-subhalos that survived with bound dark matter halo cores
($V_{\rm max} > 8$ km s$^{-1}$) experienced no stellar mass loss.  Indeed,  93\% of the surviving dwarf
galaxies in the \citet{BJ05} simulations lost less than 10\% of their initial stellar material.  The only surviving 
systems that show any stellar mass loss are those that have lost more 
than 90\% of their initial dark matter mass.  Stellar mass loss
becomes typical (affecting more than 50\% of systems) only in the minority of subhalos that have lost more than 95\% of their
initial dark matter mass.   Moreover, we find no significant
trend between current Galacto-centric distance and stellar mass loss in the simulated satellites.  These facts 
provide encouraging support to the simple assumptions we adopt here.  

Once $L$ is determined for each subhalo, we assign a stellar velocity dispersion by adopting the
 empirical relation shown in the right panel of Figure 2:
\begin{equation}
 \sigma_\star = 6.9 \, \kms \, \left(\frac{L}{10^5 \, \Lsun}\right)^{0.09},
 \end{equation}
with a log normal scatter of $\Delta \log_{10} \sigma = 0.1$ at fixed L (as measured in the data).~\footnote{We 
do not account for any systematic surface brightness bias that would lead to high $\sigma_\star$ systems being missed
(as these are the systems that will have large $\Rhalf$).  
By ignoring this effect we are systematically
{\em under-estimating} the possible number of stealth galaxies.} 
Here, we are assigning stellar velocity dispersions to our subhalos at z = 0 using a relation measured for real Milky Way dSphs. This means that the assignment is reasonable for subhalos, modulo the concern discussed above regarding stellar mass loss.~\footnote{Of course, if there is stellar mass loss,
the least-bound (hottest) stars are stripped first, and this causes the velocity dispersion and luminosity to evolve
in concert, as detailed in \citet{Penarrubia08b} for several example orbits.} 

Once $\sigma_\star$ is assigned, we determine $\Rhalf \simeq 0.75 \, \rhalf$  
using $V_{NFW}(\rhalf) = \sqrt{3} \, \sigma_\star$ (Equation 2).  The $V_{\rm max}$ and $r_{\rm max}$ values that define
$V_{NFW}(r)$ are those measured for each subhalo in the simulation (at the present day, not at the time of infall).  
Specifically, we are extrapolating the density profiles of the subhalos to radii $< 300$ pc that are  not
well resolved in the simulation.  While this extrapolation is reasonable, it means that our derived $\Rhalf$ values
are reliant on this assumption.

%>>>>>>>>>>>>>>>>>>>>>>>>>>>>>>>>>Mhalo to Mstar Mappings<<<<<<<<<<<<<<<<<<<<<<<<<<<<<<<<<
\begin{figure*}[bth!]
    \includegraphics[width=0.32\textwidth]{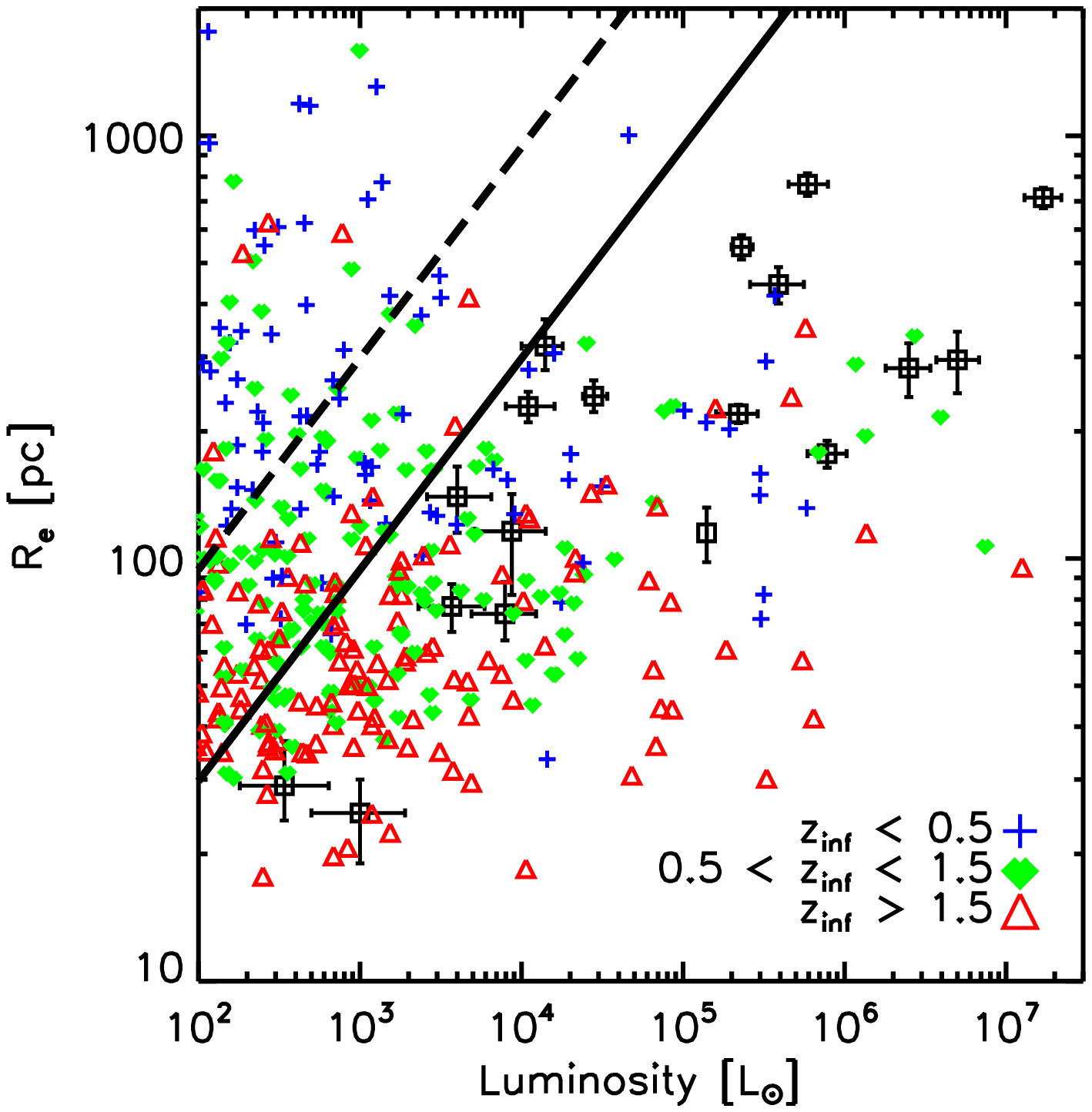}
      \includegraphics[width=0.32\textwidth]{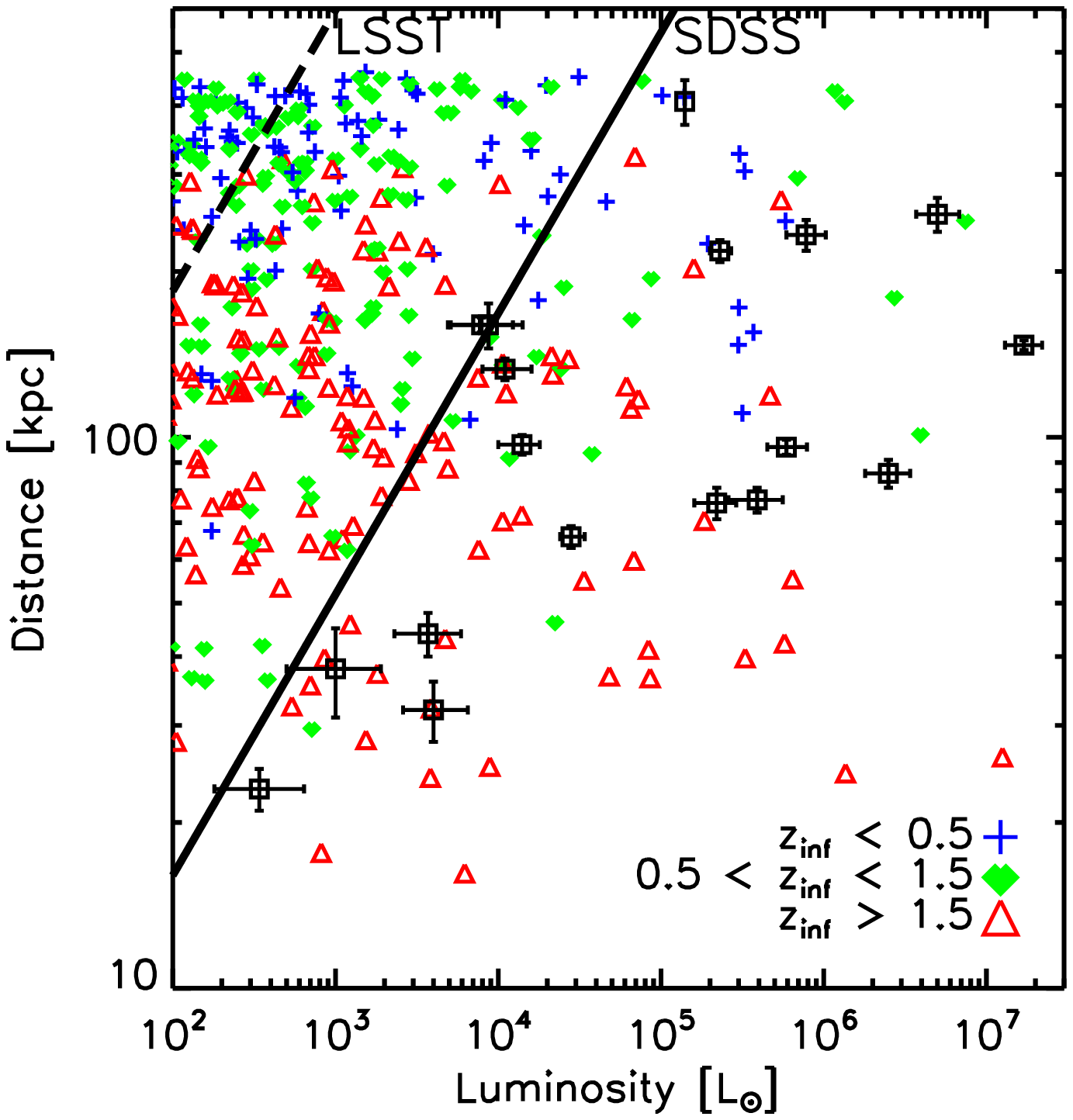}
          \includegraphics[width=0.32\textwidth]{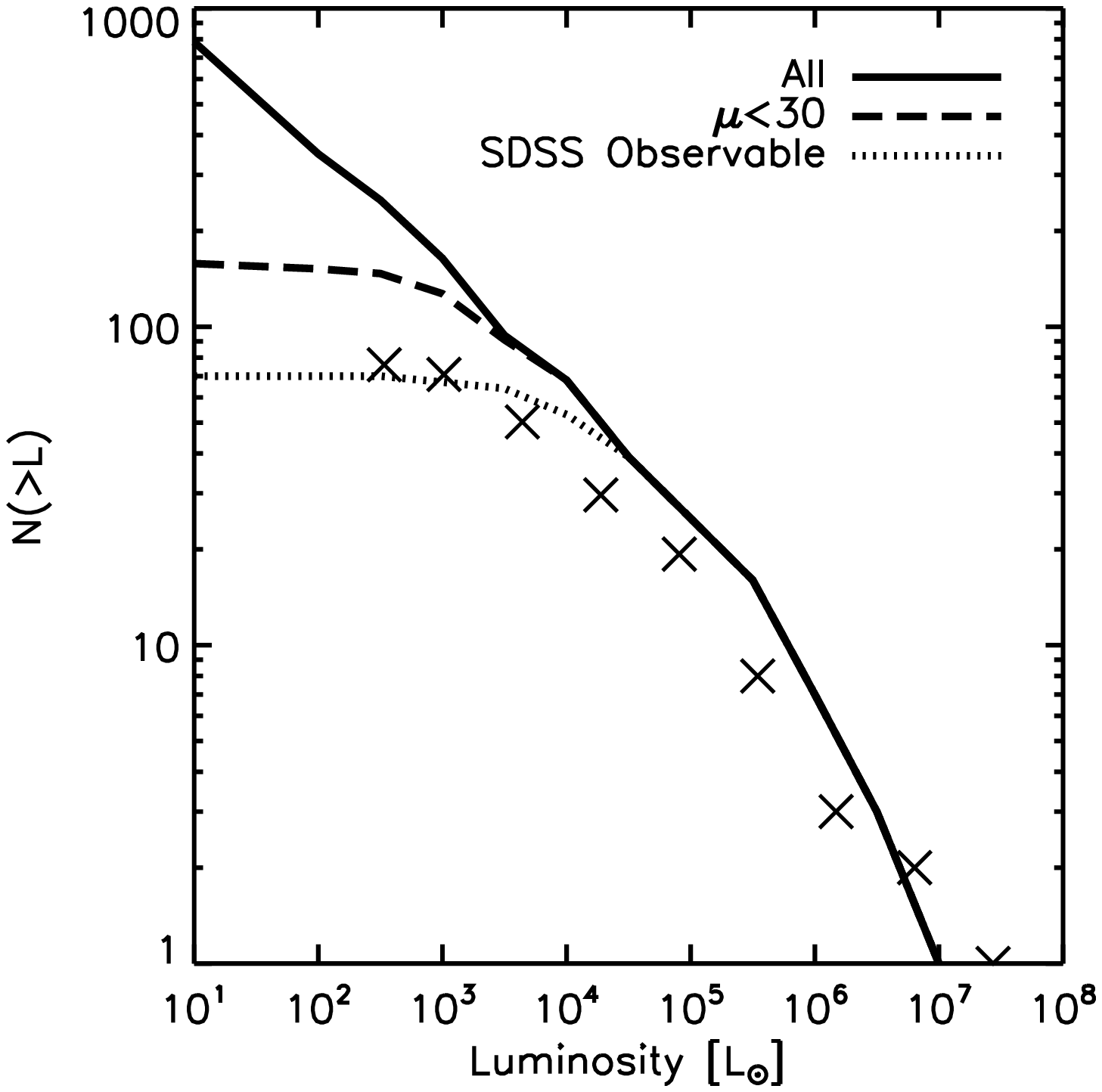}
   \caption{
   Fiducial Scenario galaxy
   size vs. luminosity relation (colored points, left); distance
   vs. luminosity relation (middle); and luminosity functions for different completeness cuts (right).  
   {\em Left Panel:} The solid line corresponds to the current detection limit at a peak surface brightness of $\mu = 30$ mag arcsec$^{-2}$ 
   and the dashed line corresponds to $\mu = 35$ mag arcsec$^{-2}$, for reference.
   The small colored points represent model galaxies and the point-type scheme maps to the redshift of infall into the
   host dark halo: open triangles have $z_{\rm inf} > 1.5$, green diamonds have $0.5 \le z_{\rm inf} \le 1.5$, and blue plusses
   were accreted since $z=0.5$.  The black squares with error bars are known MW dSph galaxies.
     {\em Middle Panel:}  The solid and dashed lines  indicate luminosity-distance completeness in the SDSS and LSST, respectively, for
   systems with $\mu < 30$ mag arcsec$^{-2}$.  The point types are the same as in the left panel.
   {\em Right Panel:}   The symbols shown as X's reflect the current census of MW dSphs, corrected for the
   sky coverage completeness of SDSS as in Tollerud et al. (2008).  The uncertainty in this correction corresponds roughly to the size of the
   symbols we use.  The dotted line shows the predicted cumulative luminosity function of satellite galaxies that are bright enough to
   have been detected by SDSS according to the Walsh et al. (2009) limits.  The dashed line shows the predicted luminosity function
   of all satellites with surface brightness meeting the $\mu < 30$ mag arcsec$^{-2}$ threshold, most of which should be
   detectable by LSST.    The solid line shows all satellite galaxies, including the stealth population.  We see that the
   majority of ultrafaint dwarfs are expected to have surface brightnesses so low that they will avoid detection without revised
   techniques for discovery.
   }
\label{fig:mappings}
\end{figure*}
%>>>>>>>>>>>>>>>>>>>>>>>>>>>>>>>>>>>>>>>><<<<<<<<<<<<<<<<<<<<<<<<<<<<<<<<<

 For simplicity, we assume that each dwarf galaxy
follows a Plummer profile, with a peak surface density given by Equation \ref{eq:plum}.
 As discussed above, galaxies with $\Sigma_{\rm peak} < 0.036 \, L_{\odot} \, {\rm pc}^{-2}$ are assumed to be undetectable with standard techniques.  
We note that if we  impose a floor in allowed velocity dispersions near $\sigma_\star = 4 ~\kms$ then our results do not change dramatically.

The final step in our procedure is a self-consistency check to see if our implied $\rhalf$ values are small enough for the galaxies to be relatively unaffected by tidal stripping.  In order to do this we estimate a tidal radius
for each galaxy $r_{\rm t}$ and remove galaxies from our catalogs if $\rhalf > r_{\rm t}$ on the assumption that most of their stars will have been tidally liberated (even though a dark matter core remains bound).  It is well known that a subhalo's rotation curve
should decline more rapidly than an NFW profile for  $r \gtrsim r_{\rm max}$ because of tidal effects \citep[][]{kazantzidis04}.  This means that $r_{\rm t} = r_{\rm max}$ provides a reasonable estimate for the gravitational tidal radius of a galaxy embedded within that subhalo.  We find that 2\% (10\%) of our galaxies with $L > 100 L_\odot$  (10 $L_\odot$) have $\rhalf > r_{\rm max}$, and that the majority of these systems have lost more than 90\% of their dark matter mass since falling into the host.  
Physically, these objects with $\rhalf > r_{\rm max}$ represent systems that are losing stellar material.
We will not explore the observational implications of this evaporating population here, but this definition may prove useful for future theoretical explorations aimed at predicting the fraction of dwarf satellites that should be showing signs of ongoing stellar stripping. By excluding these stripped galaxies with large $\rhalf$ values from our estimates we are being conservative in the sense that we are under-estimating the stealth population by 10\% for $L > 10 L_\odot$.
\citet{Penarrubia08b}  have argued that systems that are losing stellar material such as these can actually evolve towards higher mass to light ratios once the total bound mass within the stellar core decreases by $\sim 75\%$. 
  The central surface brightness typically drops by about 2 magnitudes  during the process as the stellar system expands to adjust for the loss of central mass.  Further investigation of this possibility and its implications for the stealth galaxy population is warranted, but beyond the scope of this paper.

Before moving on to our results, we mention that the  halo finder and the associated definition of halo mass
used by  Moster et al. (2009) in our stellar mass assignment differ slightly from those used in the VL2 catalogs.
 We estimate that this amounts to a $\sim 20 \%$ difference in dark matter halo mass association
 for any individual object, a difference that is not significant given the exploratory nature of this
work.

%>>>>>>>>>>>>>>>>>>>>>>>>>>>>>>>>>Moster M vs L<<<<<<<<<<<<<<<<<<<<<<<<<<<<<<<<<
\begin{figure}[tbh!]
  \includegraphics[width=0.49\textwidth]{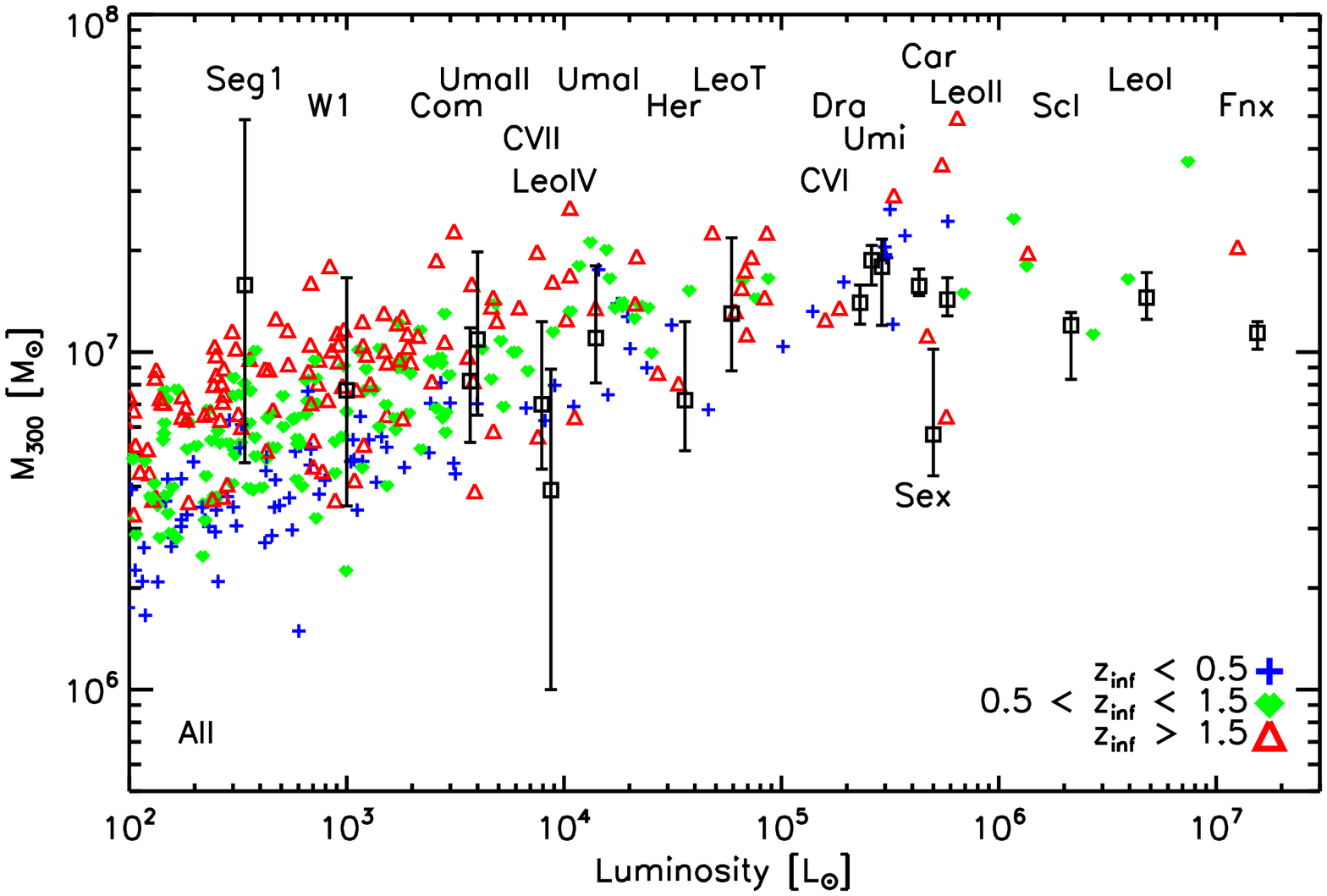}
\\
      \includegraphics[width=0.49\textwidth]{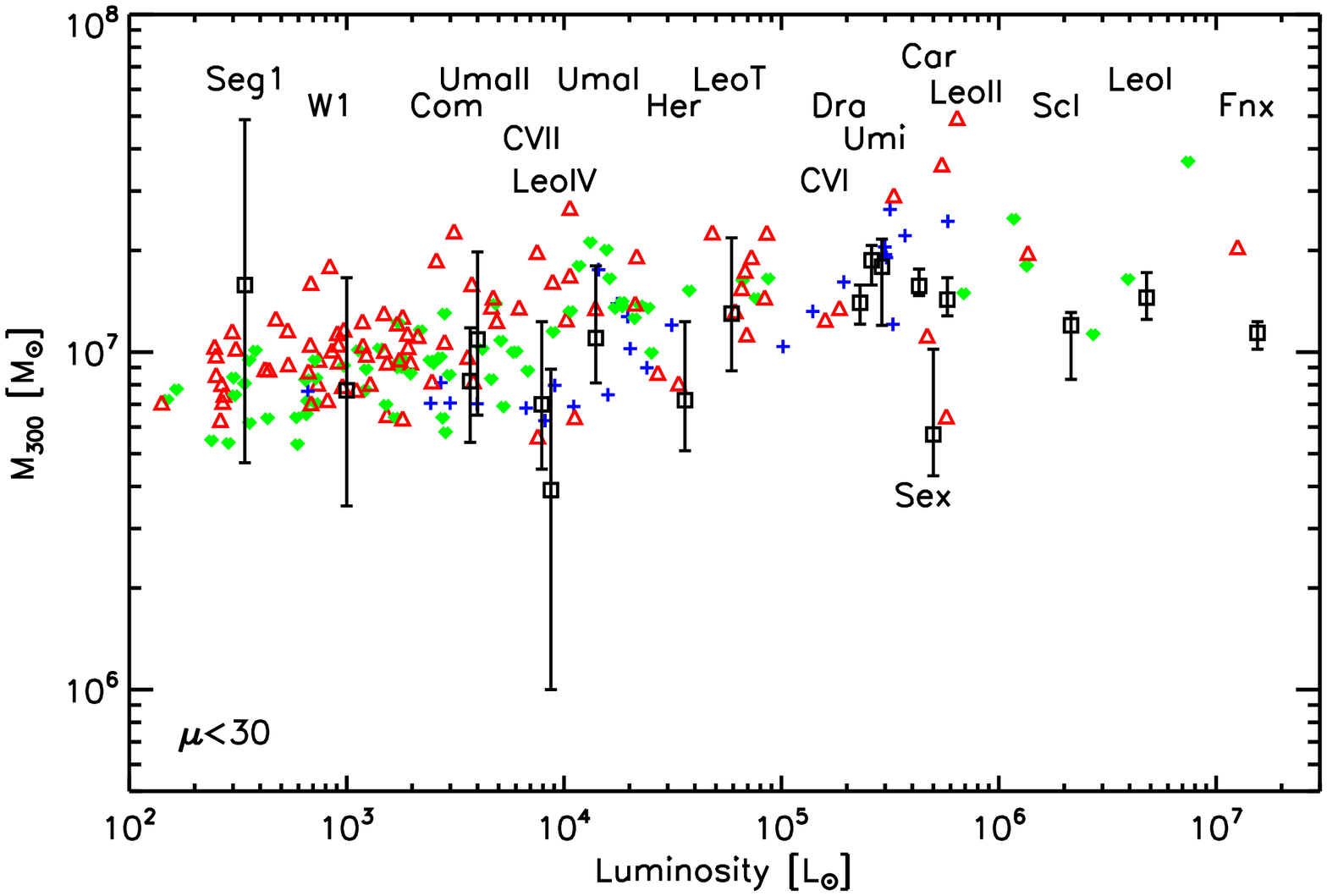}
 \\
    \includegraphics[width=0.49\textwidth]{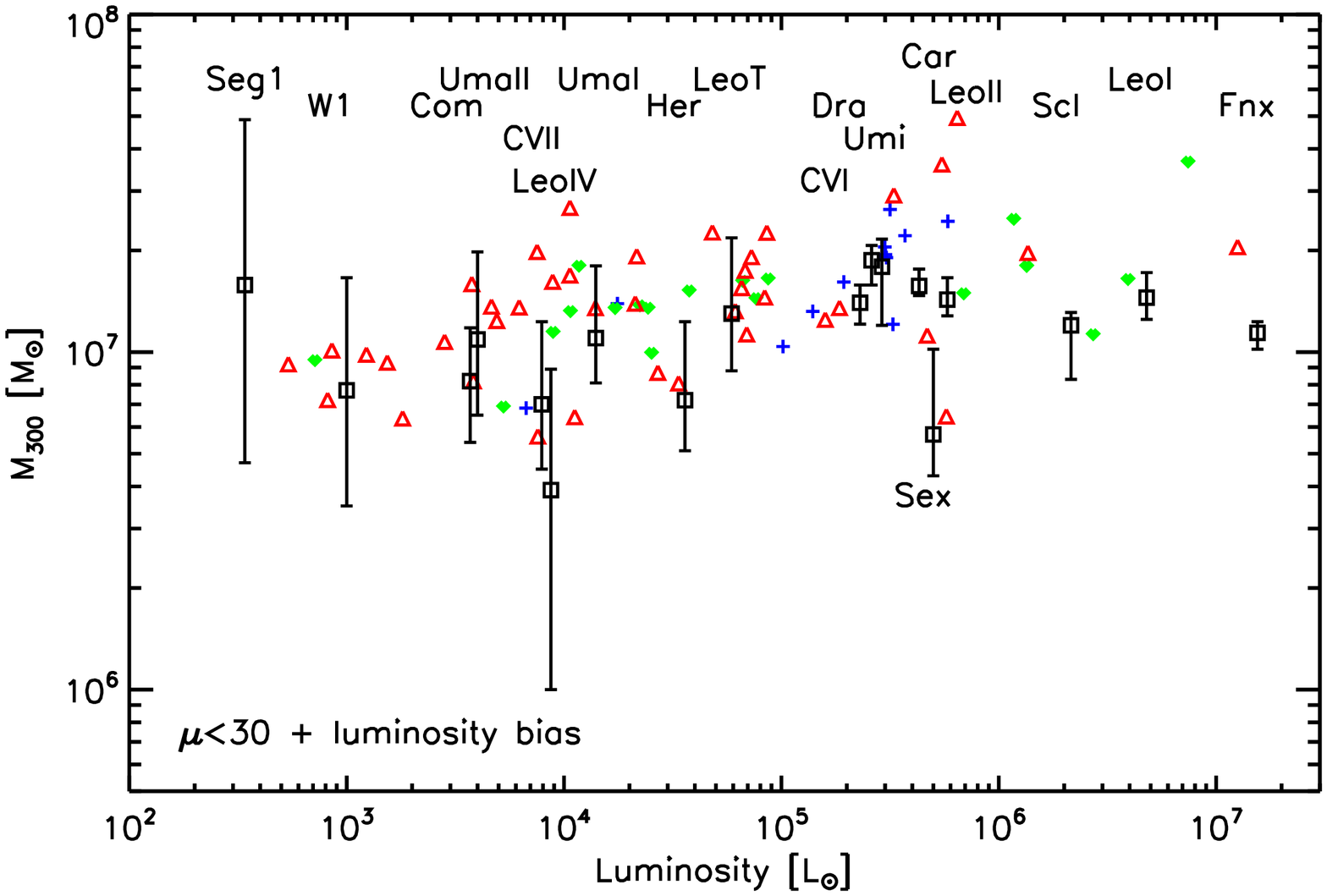}
   \caption{Mass within 300 pc as a function of luminosity.    Milky Way dSph galaxies are shown as black squares.
 Fiducial model galaxies are show as smaller colored points, with the point type and color mapped to the time
they fell in to the virial radius of the VL2 main halo.  Earlier accretions are red and recent accretions are blue as in Figure 4 and as
indicated in the upper panel legend.
The {\em upper panel} presents all predicted satellites and the {\em middle panel} shows only model satellites that are concentrated enough
to be detected with current methods, with $\mu < 30$ mag arcsec$^{-2}$.  The {\em bottom panel} includes only the subset of
$\mu <  30$ mag arcsec$^{-2}$ galaxies that are close enough to have been detected
 by SDSS, according to the 90\% completeness limits in Walsh et al. (2009). 
 The $M_{300}$ masses presented here for MW dSphs are those derived by Strigari et al. (2008).  
 We note that subsequent work has suggested that the velocity dispersion (and hence mass) errors on Hercules \citep[Herc,][]{Aden09} and W1 (B. Willman \& M. Geha, private communications) 
    are likely underestimated because of membership issues. }
\label{MvsL1}
\end{figure}
%>>>>>>>>>>>>>>>>>>>>>>>>>>>>>>>>>>>>>>>><<<<<<<<<<<<<<<<<<<<<<<<<<<<<<<<<

\section{Results}

Figure \ref{fig:mappings} provides a summary of our fiducial model predictions compared to current observations: galaxy $\Rhalf$ vs. luminosity (left), helio-centric distance vs. luminosity (middle), and 
cumulative number vs. luminosity (right).  The small colored symbols in the left and middle panels are model galaxies, with color and symbol type indicating three infall redshift bins with $z_{\rm inf} > 1.5$ (red triangles),  $0.5 \le z_{\rm inf} \le 1.5$ (green diamonds), and $z < 0.5$ (blue plusses).  The larger, black squares reproduce the MW dSph data from Figure 1. In the middle panel, the helio-centric distance for model galaxies
is measured from an arbitrary point 8 kpc from the host dark matter halo center. Our gross results are independent of this choice for solar location.  

Model galaxies above the solid line in the left panel of Figure \ref{fig:mappings} are too diffuse to have been detected.  Clearly, this population is significant.  At fixed luminosity, systems above the solid line ($\mu = 30$ mag arcsec$^{-2}$) tend to have been accreted more recently (blue plusses were accreted since $z =0.5$).  This trend follows directly from
our redshift-dependent mapping between $L$ and halo mass --  at fixed stellar mass, the required halo mass increases with redshift (Figure 3) and, as discussed above, more massive halos tend to host
more concentrated stellar distributions for a given $\sigma_\star$ and $L$.  We note that for brighter luminosities
$L \gtrsim 10^6 L_\odot$ our systems tend to be smaller in physical size than the observed MW dwarfs.  As discuss
below, this is related to the fact that they inhabit more massive dark matter subhalos ($M_{300} \simeq 3 \times 10^7$) than than appears to be the case for the MW dwarfs.
The right panel of Figure \ref{fig:mappings} reveals that 
there is a slight tendency for early-accreted galaxies (red triangles with $z_{\rm inf} > 1.5$)  to be closer to the Sun than more recently-accreted galaxies (middle panel).  This biases them to host older stellar populations.

%>>>>>>>>>>>>>>>>>>>>>>>>>>>>>>>
\begin{figure*}[tbh!]
    \includegraphics[height=0.48\textwidth]{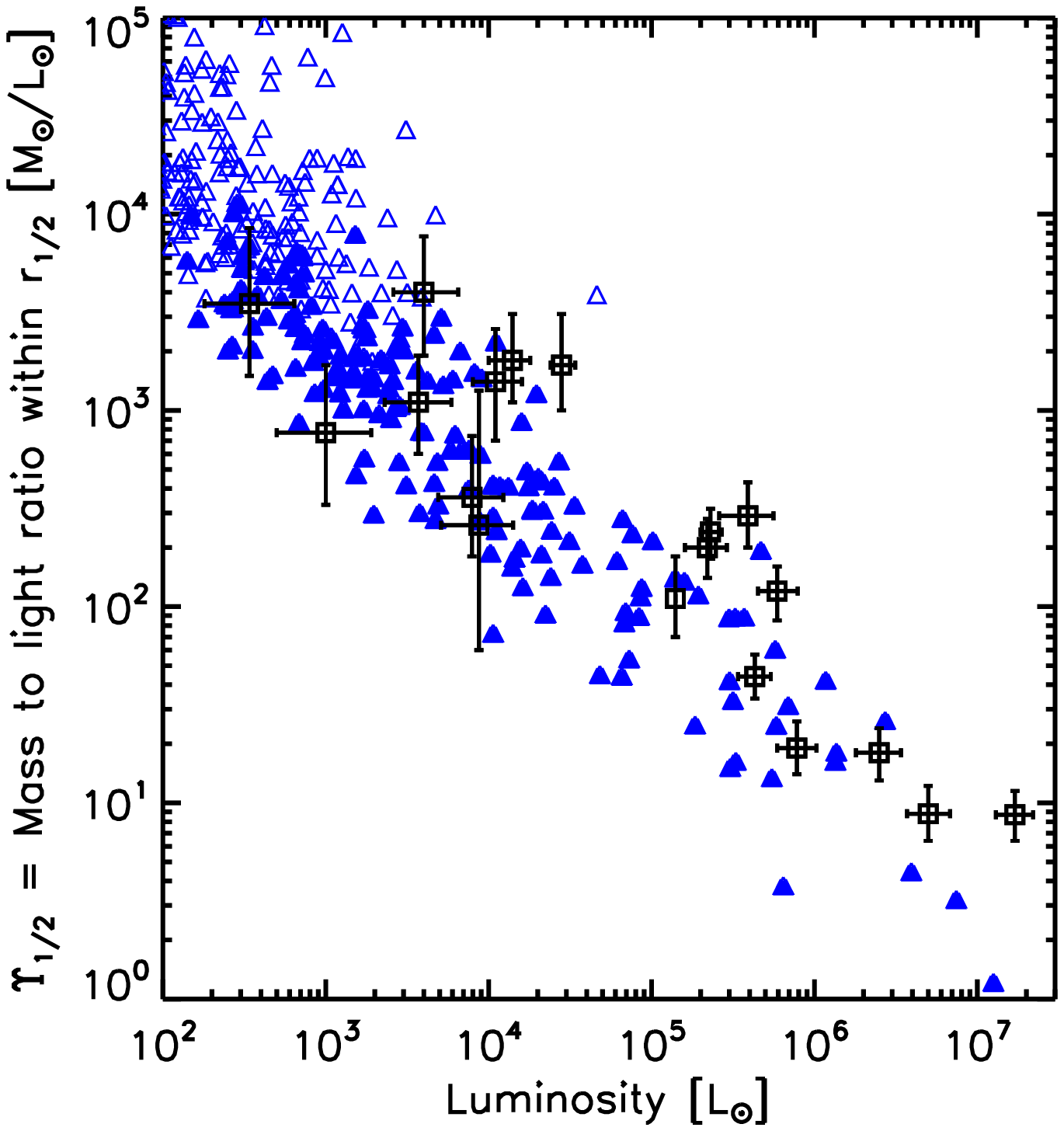}
      \includegraphics[height=0.48\textwidth]{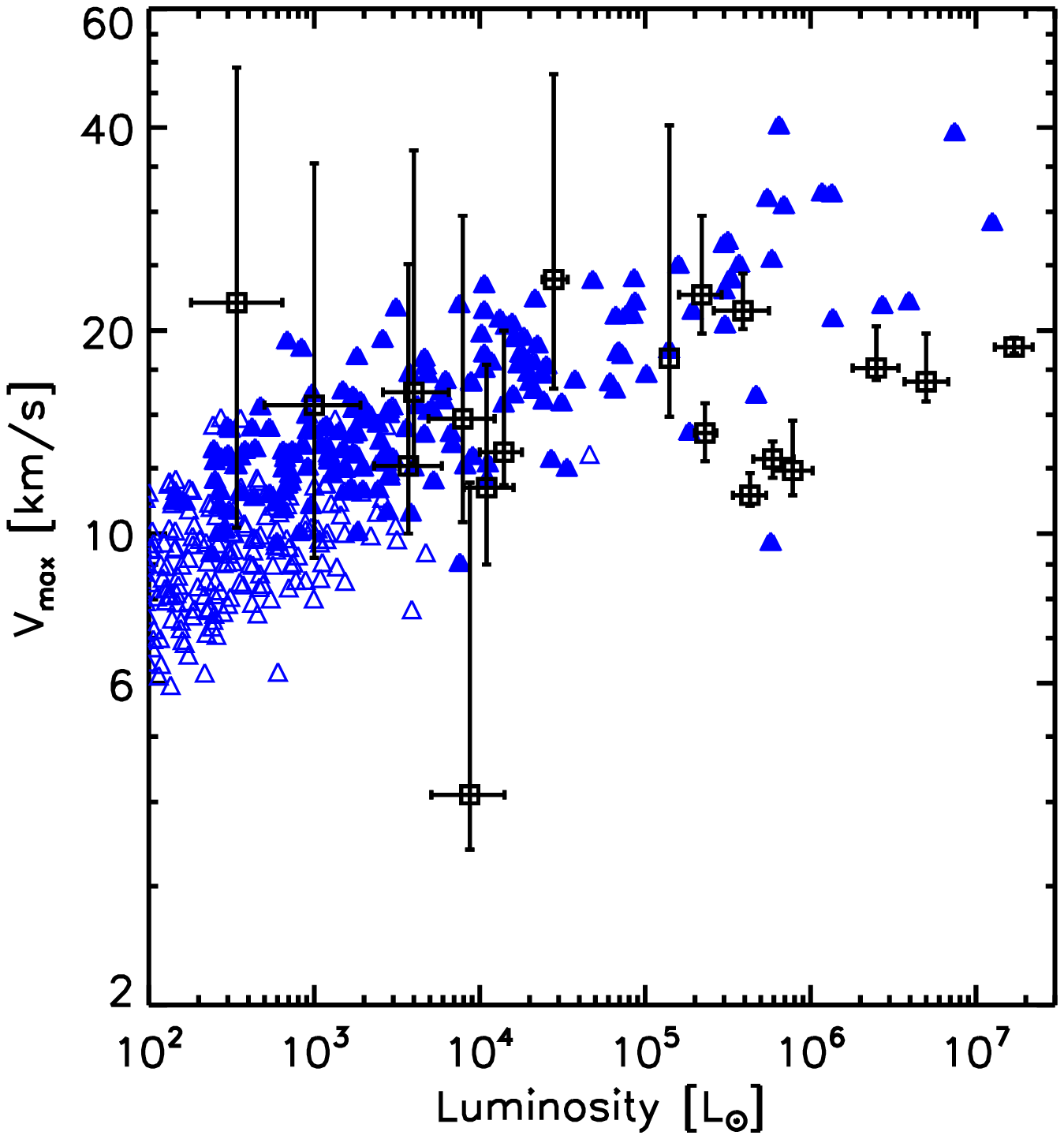}
   \caption{The dynamical mass-to-light ratio within the 3d half-light radius  vs. luminosity (left) and $V_{\rm max}$ vs. luminosity (right).
  The black squares are Milky Way dwarf spheroidal galaxies, with mass estimates are from \citet{wolf:09} and the
  $V_{\rm max}$ estimates are from Wolf et al. (in preparation) . The solid triangles show Fiducial Scenario model galaxies that are concentrated
   enough to be detected by current methods ($\mu \le 30$ mag arcsec$^{-2}$).  Open triangles show stealth galaxies
   from this same model ($\mu > 30$ mag arcsec$^{-2}$).  We see that stealth galaxies are biased to more dark matter dominated within $\rhalf$ than are their higher surface brightness counterparts.  The reason is simply that they have
   larger $\rhalf$ values.  In reality, these model galaxies reside in dark matter halos that are {\em less} massive in terms of their overall bound mass, $V_{\rm max}$, as can be seen in the right panel.
      }
\label{fig:mhalf}
\end{figure*}
%>>>>>>>>>>>>>>>>>>>>>>>>>>>>>>>>>>>>>>>><<<<<<<<<<<<<<<<<<<<<<<<<<<<<<<<<

The dotted line in the right panel of Figure \ref{fig:mappings} shows the predicted luminosity function of satellites that are observable for an SDSS-like survey covering the full sky according to the luminosity-distance completeness limits of Walsh et al. (2009). This should be compared to the data points,
which reflect the current MW satellite population corrected for sky-coverage as in \citet{Tollerud08}.  The
uncertainty in the sky-coverage correction (associated with the possibility of an anisotropic satellite distribution on the sky) is similar to the size of the points (Tollerud et al. 2008).   We see that the predicted and observed populations are roughly consistent. The solid line shows the predicted luminosity function for all satellites within 400 kpc, without any allowance for observational incompleteness.   The dashed line, on the other hand, shows the subset of those galaxies that have peak stellar surface densities 
that are bright enough to be discovered with standard techniques.  We see that roughly half of the systems that are in principle luminous enough to be detected with deep surveys like LSST (with $L \lesssim 1000 L_\odot$) have peak surface densities that are too diffuse to be seen.  Specifically, even a survey like LSST, with a very deep limiting magnitude, will have difficulties detecting these systems without new observing strategies.

Figure \ref{MvsL1} explores how detection bias affects the Strigari plot.  In the upper panel, 
we show
$M_{300}$ vs. $L$ for all model satellites within 400 kpc of the Sun (color/type scheme is the same as in Figure \ref{fig:mappings}) compared to the MW dSphs (black squares).  We see that unlike the data, there is a significant population of predicted galaxies with low central densities $M_{300} \lesssim 5 \times 10^6 M_\odot$ for $L \lesssim 5000 L_\odot$ .    Moreover, while the data follow a nearly common-mass relation for $M_{300}$ vs. $L$, the model points prefer a steeper trend:  $M_{300} \propto L^{c}$ with  $c \simeq 0.15$ (as expected from abundance matching).  The model predictions are very similar to those presented in many past CDM-based explorations of satellite $M_{300}$ values
\citep[e.g.][]{busha:09,munoz:09}.  The similarity between our model results and those of \citet{munoz:09}, in particular, are encouraging.  These authors  use the same VL2 catalog that we use, but they explored a more sophisticated model for assigning light to subhalos.
Generally, a population of $L \lesssim 5000 L_\odot$ satellite galaxies with $M_{300} \lesssim 5 \times 10^6 M_\odot$ seems to be
a  fairly robust expectation for hierarchical models, especially if $H_2$ cooling plays a role in the formation stars in surviving galaxy halos at $z=0$ \citep{munoz:09}.

 The middle panel
of Figure \ref{MvsL1} includes only those model galaxies that have peak surface brightness $\mu < 30$ mag arcsec$^{-2}$.  We see that this
requirement immediately removes the population of $M_{300} \lesssim 5 \times 10^6 M_\odot$ objects.  The lower panel includes only those
galaxies that meet both the surface brightness requirement and the luminosity-distance requirement for SDSS discovery.
We see that the resultant population of observable model galaxies has $M_{300}$ values that are very much in line with those of the known MW dSphs.
The model we have adopted therefore reproduces both the luminosity function and something close to the mass-luminosity trend seen for Milky Way dwarfs,
once all of the relevant selection bias effects are taken into account.

While we predict that stealth galaxies reside within dark matter subhalos that have the smallest total mass (as characterized 
 by $V_{\rm max}$ or $M_{300}$), this does not mean that these systems will have small dark matter fractions within their stellar extents.  On the contrary, stealth galaxies have large half-light radii for their luminosities, and therefore sample a large portion of their dark matter halos.  Indeed stealth galaxies should be {\em more} dark matter dominated within their stellar radii than higher surface brightness ultra-faint dwarfs.   It is straightforward to understand this by considering  the integrated mass within its 3d half-light radius (Wolf et al. 2010): $M(\rhalf) = 3 \, G^{-1} \,  \rhalf \, \sigma_\star^2$. For a fixed velocity dispersion (which we assume is dictated, on average, by luminosity) systems with larger half-light radii will have proportionally larger half-light masses.
Without some theoretical prior, this mass by itself tells us little about the total dark matter halo mass of the dSph host subhalo, but it does enable an easy measure of the mass-to-light ratio within the half-light radius $\Upsilon_{1/2} = M(\rhalf)/(L/2)$.  

The left panel of Figure \ref{fig:mhalf} shows $\Upsilon_{1/2}$ vs. L for the MW dwarfs \citep[open squares from][]{wolf:09} compared to
our Fiducial model galaxies (filled and open triangles).   The right panel of Figure \ref{fig:mhalf} shows the same
simulation data with the same symbol types but now with halo $V_{\rm max}$ vs. L.  The MW dwarf $V_{\rm max}$ estimates
are obtained with a CDM subhalo prior as described in Wolf et al. (in preparation).
 The solid triangles correspond to predicted galaxies that are concentrated
   enough to be detected by current methods and open triangles show predicted stealth galaxies
($\mu > 30$ mag arcsec$^{-2}$).   As expected, the stealth galaxies are biased to more dark matter dominated within $\rhalf$ than are their higher surface brightness counterparts. 
The implication is that stealth galaxies should  have integrated mass-to-light ratios upwards of $\Upsilon_{1/2} \simeq 10^4$.  Again, this is a direct result of the fact that they have large half-light radii.   As illustrated in the right panel, the same galaxies actually reside within
subhalos that are less massive overall, with $V_{\rm max}$ values that are systematically smaller than $\sim 10$ km s$^{-1}$.
It is important to remind the reader that 
the precise values of $M(\rhalf)$ for $\rhalf \lesssim 300$ pc in the theory points are derived by extrapolating the subhalo profiles inward to radii that are not well-resolved in the simulation.  On the other hand, the $V_{\rm max}$ values for the data are extrapolated
outward from $\rhalf$.  So while neither comparison is ideal,  the approximations are reasonable, and
the qualitative behavior that distinguishes stealth galaxies from more readily observable galaxies is expected to be robust.
Note that the predicted galaxies at the bright end of the luminosity distribution are both too small (with lower $\Upsilon_{1/2}$ values
than the data in the left panel) and too massive (higher $V_{\rm max}$ values than the data in the right panel).  This is the
same problem discussed above in association with Figure 4, and stems from the fact that the VL2 simulation contains several
subhalos that are more massive than any of the dSph satellites of the Milky Way.

%>>>>>>>>>>>>>>>>>>>>>>>>>>>>>>>
\begin{figure*}[tbh!]
    \includegraphics[height=0.4\textwidth]{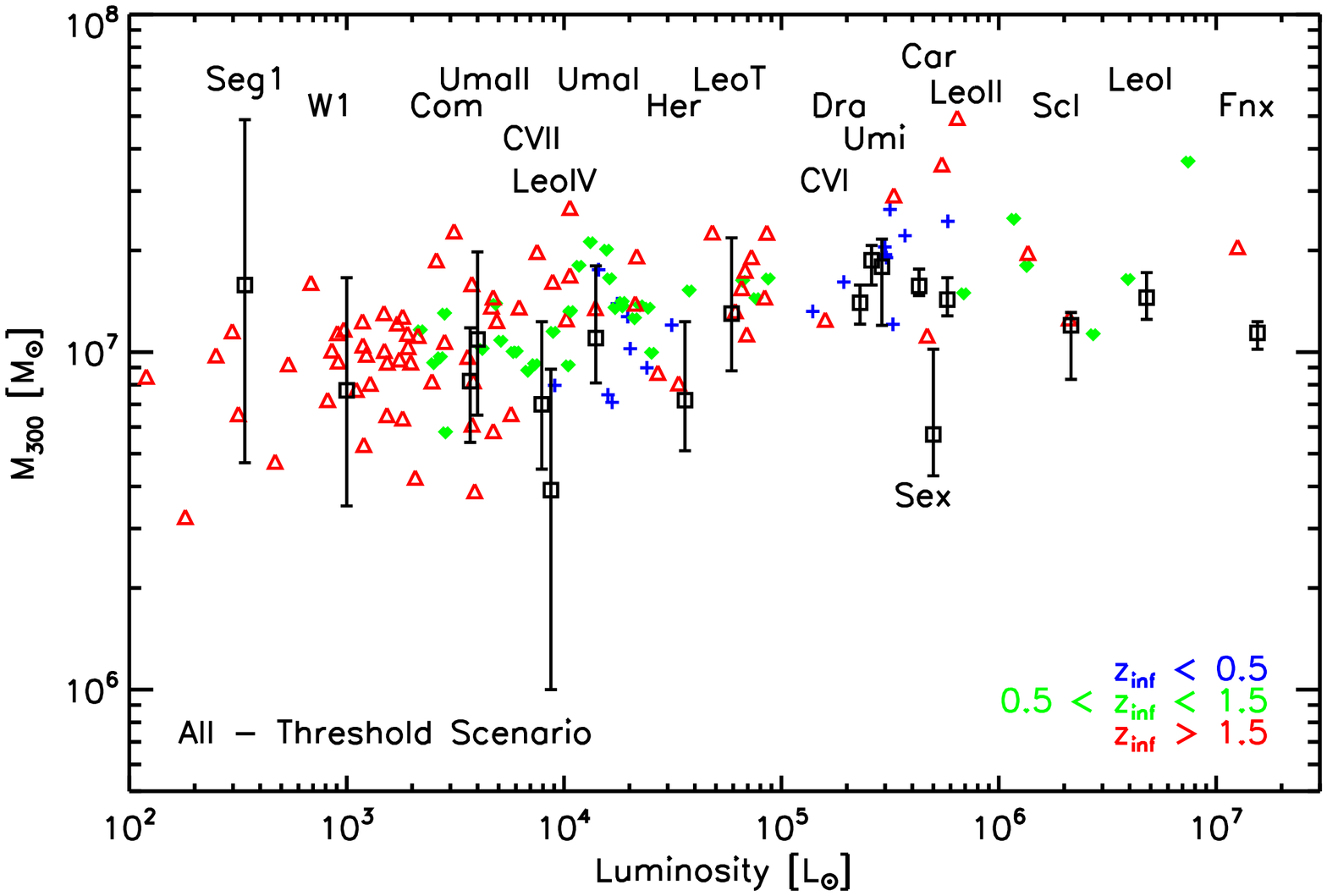}
  \includegraphics[height=0.4\textwidth]{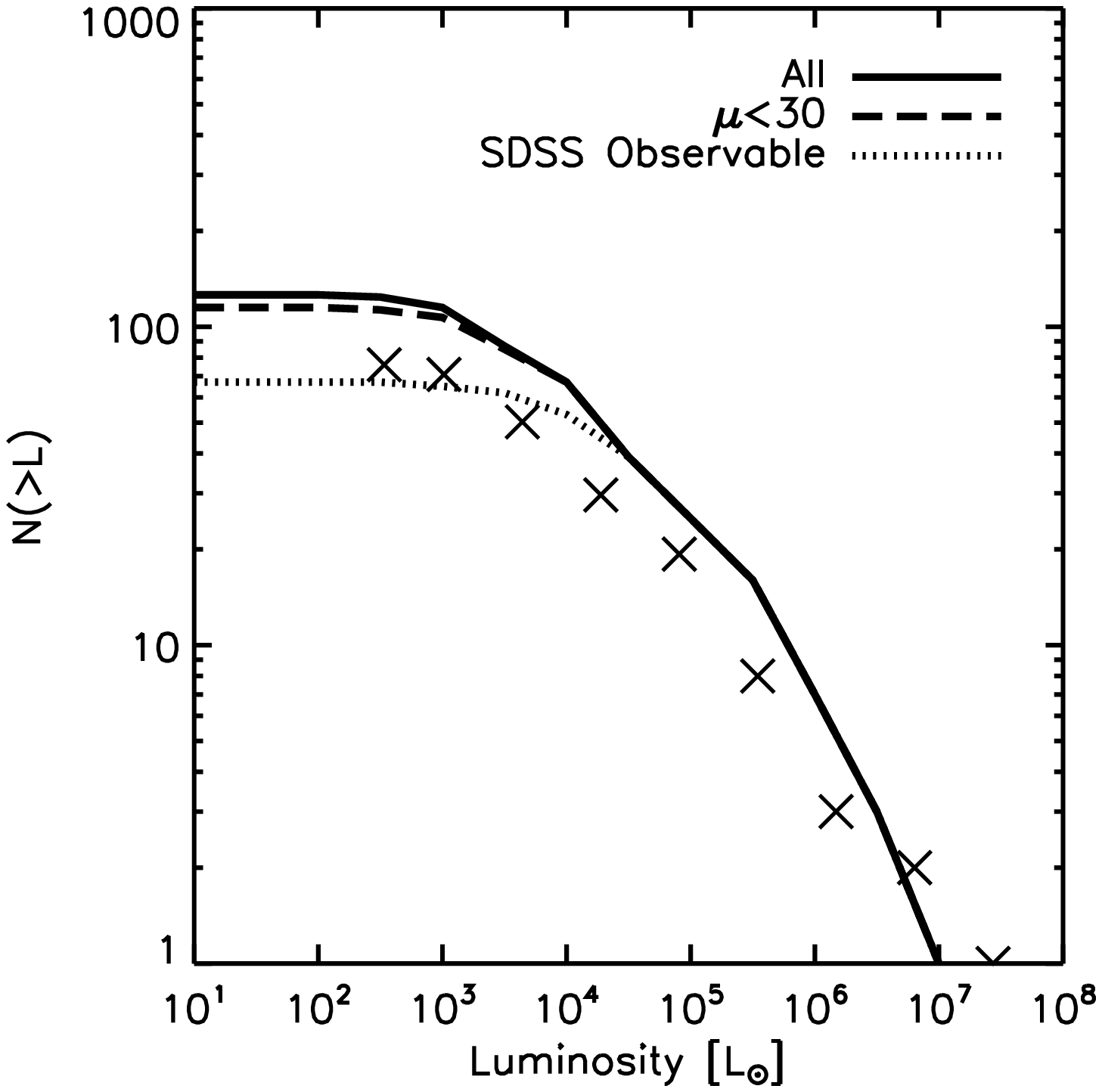}
   \caption{Threshold Scenario predictions for the Strigari plot (left) and luminosity function (right).  The symbols and line types are identical to those
   in Figure 5 and the right panel of Figure 4, respectively.   This model imposes a sharp truncation in galaxy formation efficiency at $\Mvir = 5 \times 10^8 \Msun$, which drastically reduces the expected number of stealth galaxies.
      }
\label{fig:mappings2}
\end{figure*}
%>>>>>>>>>>>>>>>>>>>>>>>>>>>>>>>>>>>>>>>><<<<<<<<<<<<<<<<<<<<<<<<<<<<<<<<<

Of course, if the smallest dark matter halos do not contain galaxies at all, then the likelihood for a significant stealth galaxy population is much reduced.  
We explore this expectation with our Threshold Scenario, which imposes a sharp scale in galaxy formation
at $\Mvir = 5 \times 10^8 \Msun$.  Below this scale dark matter halos are completely devoid of stars (vertical line in Figure 3).
The resultant Strigari plot and luminosity functions for this model are shown in  Figure \ref{fig:mappings2}.
 Like the Fiducial Scenario,
the Threshold Scenario also reproduces the observed satellite luminosity function (right panel).  However,
unlike in the Fiducial Scenario,
we now expect only a handful of stealth galaxies that remain undiscovered (solid vs. dashed lines).
The Threshold case also yields a Strigari relation that is in reasonable agreement with the data (left panel), without appealing to any
selection bias.  Another distinct difference in the Threshold Scenario is that
 all of the low-luminosity galaxies are expected to be quite old (or at least to have been accreted
early $z_{inf} > 1.5$).

One problem with both of our models is that they produce too many massive dwarf satellites with $M_{300} \gtrsim 2 \times 10^7 M_\odot$ 
compared to the data (Figures \ref{MvsL1}, \ref{fig:mhalf}, and \ref{fig:mappings2}).  This has nothing to do with our
method of light assignment.  There are simply too many massive subhalos in the VL2 halo when compared to the MW satellite population.
The same problem is responsible for the fact that our model galaxies tend to be too small at $L \gtrsim 10^5 L_\odot$ compared to the data in the left panel of Figure \ref{fig:mappings}.
These model galaxies reside in subhalos that are more massive than the subhalos that host the brightest galaxies in the Milky Way, and this confines their stars to a 
characteristically smaller radius.  Of course, this issue may simply reflect cosmic variance in the subhalo populations from galaxy to galaxy, but it could be attributed to the fact that VL2 represents a slightly more massive halo than the Milky Way's dark
matter halo.

It is important to mention that the $M_{300}$ masses in Figures 2, \ref{MvsL1} and \ref{fig:mappings2} 
are taken directly from the spherical Jeans modeling of
Strigari et al. (2008).  A more recent analysis of membership in the Hercules dwarf (labeled Her in Figures \ref{MvsL1} and \ref{fig:mappings2})
by \citet{Aden09} suggests that the actual $M_{300}$ mass for this system is about  a factor of $\sim 2$ lower than the value we have used.  
Though this result does not change
the fact that there is no strong observed trend between $M_{300}$ and $L$ in the data, it does make the mass of Hercules more difficult to explain
in our model (and in almost all CDM-based models to date).  In the right panel of Figure 2, the \cite{Aden09} velocity dispersion for Hercules
$\sigma_\star \simeq 3.7 \, \kms$ would shift the point at $L \simeq 3 \times 10^4 L_\odot$ down to the edge of its error-bar --
 clearly low enough that this system would not be expected to be stealth according to our definition.  
The fact that the velocity dispersion shifts this much by removing or adding a few stars suggests that there is a large membership-related
systematic error that needs to be taken into account in the mass modeling.~\footnote{This is not the only systematic uncertainty.  As the most elongated of the 
dSphs known, Hercules is not particularly well suited for spherical Jeans modeling in the first place.}
More work on the issue of membership and mass modeling in this interesting object is certainly warranted.

\section{Conclusions and Discussion}
\label{Conclusion}

We have argued that there is likely a population of low-luminosity satellite galaxies orbiting within the halo of the Milky Way that are too diffuse to have been detected with current star-count surveys, despite the fact that they have luminosities similar to those of known ultrafaint MW dSphs.  
These stealth galaxies should preferentially inhabit the smallest dark matter subhalos that host stars ($V_{\rm max} \lesssim 15 ~ \kms$).
  One implication is that selection bias (Figures 2 and 5) may play a role explaining the apparent common mass scale for MW dSph galaxies \citep{Strigari08,penarrubia08,walker:09,wolf:09}.  It also implies that searches for the lowest mass `fossil' galaxies left over from reionization may be hindered by surface brightness limits.  This latter point was
 made earlier by  \citet{Bovill_Ricotti2009}. 
  According to our fiducial estimates, potentially half of several hundred satellite galaxies that could be observable by surveys like LSST are stealth (Figure 4).  

We developed a plausible estimate for the number and character of MW stealth satellites using the
subhalo catalogs of the VL2 simulation \citep{VL2}.  We assigned light to subhalos by extrapolating the dark matter 
 mass-light relationship required to reproduce bright  galaxy number counts \citep{Moster09} and we assigned stellar 
 velocity dispersions to each system by adopting the empirical relationship between $\sigma_\star$ and $L$ for known Milky Way dwarfs. Finally,   galaxy sizes were computed using the dynamical relationship between
 $\Rhalf$ and $\sigma_\star$ for the measured dark matter halo densities in each subhalo (Equation \ref{eq:joewolf}).  The resultant model galaxy population includes a substantial
 fraction of ultrafaint galaxies that are stealth, with peak surface brightness $\mu > 30$ mag arcsec$^{-2}$. 
 
We also explored the possibility that there is a sharp threshold in galaxy formation at a halo mass of $\Mvir = 5 \times 10^8 \Msun$.  This idea follows from the common mass conjecture in \citet{Strigari08} and remains viable since it reproduces the observed Milky Way satellite luminosity function as well as a fairly weak $M_{300}$ vs. $L$ trend  without appealing to selection bias (Figure 6).  In this scenario, all satellite galaxies are born within halos that are quite dense, and therefore the number of predicted stealth galaxies (which preferentially inhabit the smallest dark matter halos) is significantly reduced (Figure 6).   Moreover, we expect that all of the low-luminosity satellites will have been accreted since $z \simeq 1.5$ (Figure 6) and that they will all host old stellar populations.  This is not necessarily the case in our Fiducial Scenario, where the most distant and low-mass subhalos may host ultrafaint galaxies that contain intermediate-age stars.  

A significant shortcoming in our approach is the complete lack of accounting for the effect of baryons on the density structure, distribution, and number counts of dwarf galaxy subhalos.  The VL2 simulation that we have used is purely dissipationless, and
therefore lacks a central disk component, which would act to enhance subhalo mass loss and selectively eliminate subhalos
that cross near the disk region \citep{Donghia10,Pen10a}.  At the same time, it also lacks the inclusion of any gasdynamical
effects such as adiabatic contraction, which can increase the central densities of subhalos with stars.  In fact,
\citet{WS10} found that subhalo $V_{\rm max}$ values tend to increase for the most luminous satellites ($L \gtrsim 10^6 L_\odot$)
in cosmological simulations.  Their simulations  self-consistently form a central galaxy and produce a satellite population that matches
the bright end of the luminosity function of Milky Way dSphs, so the direction of the effect needs to be taken seriously.  
However, the degree to which baryons can enhance the densities of massive subhalos is uncertain and model specific,
as certain wind models can also act to evacuate mass
from the centers of dwarf galaxies \citep{Gov10}.  All of these concerns emphasize the need for further work in the attempt
to make sense of observed dwarf galaxy structural properties.  At this time it is difficult to know  how baryons would affect our
 estimates for stealth galaxy counts.  For the faint galaxies of concern, there is very little baryonic
material left  to have driven contraction, while significant central mass loss from winds or global dark matter mass loss
from tides are both possible.  For systems that remain bound, these effects will increase the likelihood for them to be 
low surface brightness, as galaxies will tend to expand as their global dark matter potential depth decreases.
On the other hand, mass loss also enhances the rate of disruption, and can deplete the overall number of satellites.  
Given that our current model seems to reproduce the observed luminosity function (and mass function) fairly
well, one could argue that depletion has not been too much of an issue.

It is well known that galaxy formation has a primary  scale -- the scale that gives rise to the $L_*$ cutoff in galaxy counts at the bright end of the luminosity function.  We do not know if there is a second scale that operates at the low-luminosity end.  One implication of our findings is that a complete search for very low surface brightness satellite galaxies of the Milky Way can help determine whether or not there is a second scale in galaxy formation.  If very few stealth galaxies are discovered, this
will be an indication that there are no very low mass halos that host stars.  A similar effect would be seen if there were simply a truncation in the power spectrum,
as might be expected in $\sim 1$ KeV WDM models \citep{Strigari08,mf09}.  In this sense, the discovery of many stealth galaxies in the halo would provide a means to constrain dark matter particle properties in addition to galaxy formation physics.

\cite{Kollmeier09} have performed kinematic follow-up observations of the Pices Overdensity (also called Structure J) 
  at a distance of $\sim 85$ kpc and have argued that it may be a very low surface brightness dwarf galaxy.
  If this is true then it represents the first detection of a stealth dwarf galaxy in the halo of the Milky Way.
  Unlike the well-known ultrafaint dSphs of the Milky Way, which were discovered as overdensities in RGB or MS turnoff stars, 
this system was discovered as an excess in RR Lyrae variables in the multi-epoch SDSS Stripe-82 \citep{Watkins09,Sesar07}.   
 We caution that one potential problem with the stealth-galaxy interpretation of the Pices Overdensity comes from Sharma et al. (in preparation), who have used 2MASS data 
to show that this structure is consistent with being part of a larger overdensity of stars, in which case it is unlikely to be a bound dwarf. 
Deeper, wide-field imaging and spectroscopic follow-up will be required to determine the nature of  this interesting structure.  
  
It is reassuring to note that upcoming deep, time-resolved
 sky surveys and associated follow-up campaigns with 30m-class telescopes
 offer significant hope for the discovery of hundreds of new dwarf galaxy companions of the Milky Way
(Tollerud et al. 2008).   Repeated sky surveys like Pan-STARRS and LSST may provide the best hope for discoveries 
in the future. Not only will they allow the identification of variable tracers, but they will also enable concurrent
use of bulk proper motions.    Confident searches within these data offer a means to limit the overall count of stealth galaxies 
that lurk at very low surface brightness and to provide unparalleled constraints on the
efficiency of galaxy formation in the smallest dark matter halos.

\acknowledgements  
 We thank B. Barton, C. Rockosi, M. Geha, G. Gilmore, A. Kravtsov, J. Simon, M. Ricotti, L. Strigari, M. Rocha, and B. Willman for
enlightening discussions.
M. Geha, J. Simon, L. Strigari, and an anonymous referee provided valuable advice on the manuscript. 
This work was supported by
the Center for Cosmology at the University of California, Irvine.

\bibliography{stealth5}

\end{document}